%% file: main.tex
\documentclass[11pt]{article}

\usepackage{a4wide,amsmath,afterpage,rotating,latexsym,psfrag,epsf,epsfig,amssymb,float}
\usepackage{color}
\usepackage{parskip}
\usepackage{natbib}
\usepackage{url}
\usepackage{graphicx}
\usepackage{subcaption}
\usepackage{multirow}  % Required for merging rows
\usepackage{array}     % For better array customization
\usepackage{float}
\usepackage{comment}
\usepackage{authblk}
\providecommand{\keywords}[1]
{
  \small	
  \textbf{\textit{Keywords---}} #1
}

\usepackage{appendix}

\newcommand{\bfx}{\hbox{\boldmath$x$}}

\begin{document}
\title{Logistic regression models: practical induced prior specification}
\author[1,2]{Ken B. Newman}
\author[3,4]{Cristiano Villa}
\author[2]{Ruth King}
\affil[1]{Biomathematics \& Statistics Scotland, Edinburgh, UK.}
\affil[2]{School of Mathematics and Maxwell Institute for Mathematical Sciences, University of Edinburgh, Edinburgh, UK.}
\affil[3]{Division of Natural and Applied Sciences, Duke-Kunshan University, Kunshan, China}
\affil[4]{School of Mathematics, Statistics and Physics, Newcastle University, Newcastle-upon-Tyne, UK}

\maketitle

%-----------------------------------------
\input{0_Abstract}

%-----------------------------------------
\input{1_Introduction}

 %-----------------------------------------
\input{2_Logistic_Reg_Constant_Mean}

%-----------------------------------------
\input{3_Mean_Variance_Prior_Logistic_Reg_Coef}

%-----------------------------------------
\input{4_Arbitrary_beta_theta}

%-----------------------------------------
\input{5_Simulations_0}

%-----------------------------------------
\input{6_Application}

%-----------------------------------------
 \input{7_Discussion_Conclusions}

\paragraph{Acknowledgements.} 
For the purpose of open access, the authors have applied a Creative Commons Attribution (CC BY) license to any Author Accepted Manuscript version arising from this submission.
%-----------------------------------------
\bibliographystyle{apalike}
\bibliography{main}

%-----------------------------------------
 \newpage 
\appendix
\renewcommand{\appendixname}{Appendix}
\input{Appendix}

\end{document}

%% file: 0_Abstract.tex
\begin{abstract}
Bayesian inference for statistical models with a hierarchical structure is often characterized by specification of priors for parameters at different levels of the hierarchy. When higher level parameters are functions of the lower level parameters, specifying a prior on the lower level parameters leads to induced priors on the higher level parameters.  
However, what are deemed uninformative priors for lower level parameters can induce strikingly non-vague priors for higher level parameters. Depending on the sample size and specific model parameterization, these priors can then have unintended effects on the posterior distribution of the higher level parameters. % These unintended effects can be apparent from alternative parameterizations at the lower level that yield mathematically equivalent higher level parameters. 

Here we focus on Bayesian inference for the Bernoulli distribution parameter $\theta$ which is modeled as a function of covariates via a logistic regression, where the coefficients are the lower level parameters for which priors are specified. A specific area of interest and application is the modeling of survival probabilities in capture-recapture data and occupancy and detection probabilities in presence-absence data. In particular we propose alternative priors for the coefficients that yield specific induced priors for $\theta$. We address three induced prior cases. The simplest is when the induced prior for $\theta$ is Uniform(0,1). 
%Solutions have been proposed previously for the case of a single covariate but not multiple covariates. 
The second case is when the induced prior for $\theta$ is an arbitrary Beta($\alpha$, $\beta$) distribution.  The third case is one where the intercept in the logistic model is to be treated distinct from the partial slope coefficients; e.g.,    $E[\theta]$ equals a specified value on (0,1) when all covariates equal 0. Simulation studies were carried out to evaluate performance of these priors and the methods were applied to a real presence/absence data set and occupancy modelling.
\end{abstract}

\keywords{Bayesian inference, Induced priors, Logistic regression, Objective priors, Occupancy modeling.}

%% file: 1_Introduction.tex
\section{\label{sec:Introduction}Introduction}

Many processes with binary outcomes, say 0 or 1, are modeled as Bernoulli random variables $Y$ where $\Pr(Y=1) = \theta$ and $\Pr(Y=0)=1-\theta$. The value $Y=1$ can equate to a variety of nominal variables such as success,  survival or presence while the value $Y=0$ corresponds to the opposite: failure, death or absence, respectively.  Such Bernoulli processes appear in several areas of statistical ecology such as modeling the probability of an animal surviving over some time interval, the probability of an animal producing young, the probability of animal's or plant's presence at a given location and the probability of detecting such presence. 
 
 Often the Bernoulli probabilities are modelled as functions of covariates, $x_1,\ldots,x_p$, and to ensure that the probability $\theta$ satisfies the condition $0 < \theta < 1$, a logistic transformation of $\theta$ is often used:
 \begin{align*}
\ln\left ( \frac{\theta}{1-\theta} \right ) &= \beta_0 + \beta_1 x_1 + \ldots + \beta_p x_p.
 \end{align*}
 We label the logistic transformation above $logit(\theta)$. When Bayesian inference is carried out, prior distributions are specified for the coefficients $\beta_0, \beta_1, \ldots, \beta_p$. These prior distributions \textit{induce} a prior distribution on $\theta$.  

 The choice of priors for Bayesian inference and their effect has been long and frequently discussed \citep{2014_Gelman_etal, 2015_Hobbs_Hooten}. Several approaches have developed for selecting and formulating priors \citep{Jeffreys, Bernardo1979, DATTA200591,1996_Kass_Wasserman,2017_Simpson_etal,Leisenetal2020,2021_Walker_Villa,Villalobosetal2024} with advice on approaches to prior selection, e.g., \citet{2020_Banner_etal} and \cite{2021_deSchoot_etal}. Concerns about the choice of priors and criticism of Bayesian methods in general have been voiced by many, particularly in ecological applications \citep{2009_Lele_Dennis, 2020_Lele} and the use of so-called default priors, those built into various computer software, are a general concern \citep{2020_Banner_etal}. 
 
 A specific concern is that the use of seemingly innocuous ``non-informative'' priors at one level of a hierarchical model can induce strikingly non-innocuous, i.e., relatively concentrated, priors at a higher level. This has been noted specifically for the priors for the $\beta$'s in the logistic regression models \citep{2012_Seaman_etal, 2018_Northrup_Gerber, 2020_Lele}.  For example, \citet{2012_Seaman_etal} show how Normal(0,$25^2$) priors for $\alpha$ and $\beta$ in the simplest logistic model 
 $logit(\theta) = \alpha + \beta x$ induces a so-called bathtub shaped prior on $\theta$ with most of the probability mass concentrated at the extremes 0 and 1. Similarly, \citet{2018_Northrup_Gerber} show the effects of a variety of such priors in the analysis of presence/absence data and occupancy modeling where there are two Bernoulli probabilities, one for presence, and another one for detection conditional on presence, which are both often modeled as functions of covariates via logistic regression.  In the context of occupancy modeling \citet{2020_Lele} examined posterior inference for a function of the presence and the detection probabilities, namely the
probability that a site is occupied but it is observed
to be unoccupied on all visits. He pointed out sizable problems of invariance in the posterior distribution of this function to two alternative, but mathematically equivalent, parameterizations of the occupancy model, with samples of size at or around 30, where for one parameterization the point estimate is 0.30 and for the other parameterization is 0.67.  
 
 While sometimes the sample size is sufficient to yield a posterior distribution for $\theta$ that is dominated by the data and therefore relatively unaffected by the prior, the practitioner might prefer to start with induced priors that more closely match their opinion, such as $\theta$ $\sim$ Uniform(0,1) or more generally $\theta$ $\sim$ Beta($\alpha$, $\beta$).  
 
 We do note that there are, of course, cases where informative priors are desired which reflect expert opinion \citep{2008_OHagan, 2019_OHagan} and that the use of subjective priors can be defensible \citep{2020_Banner_etal}. 
 
  The purpose of this paper is to present procedures for specifying priors for the coefficients of a logistic regression model with $p$ covariates which induce priors for a Bernoulli probability $\theta$ that approximate a specified distribution, in particular those involving arbitrary Beta distributions. This includes the specific case of Beta(1,1) or Uniform(0,1). While \citet{2018_Northrup_Gerber} provide some advice for specifying priors for an intercept alone model, namely $p$=0, they do not address the effects of multiple covariates on the induced prior for $\theta$. We also consider the setting where the intercept is to be treated differently than the coefficients for the covariates; e.g.,  $E[\theta]$ equals a specified value on (0,1) when all covariates equal 0.

  In Section \ref{sec:simple.logistic} we begin with the simplest case of an intercept only model, $logit(\theta)=\beta$. We derive the induced prior for the Bernoulli probability $\theta$ given the prior for the intercept, $\pi(\beta)$, and vice versa, and present a normal distribution prior for $\beta$ that yields approximately a Uniform(0,1) for $\theta$.  Section \ref{sec:mean.var.determination} extends the approach to $p\ge$ 1 covariates to induce a Uniform(0,1) prior on $\theta$. In Section \ref{sec:arbitrary.beta} the case where the induced prior for $\theta$ is any Beta distribution is addressed along with an approach for distinct handling of the intercept. Simulation analyses are presented in Section \ref{sec:simulations} and the method is applied to a real occupancy data set in Section \ref{sec:application}.

%% file: 2_Logistic_Reg_Constant_Mean.tex
\section{\label{sec:simple.logistic}Logistic regression with constant mean model}
We focus on logistic regression where the logit
link function is used to model the probability
of ``success'' with a Bernoulli random variable:
\begin{align}
%\label{eq:obs.model}
Y &\sim \mbox{Bernoulli} \left ( \theta \right ); \nonumber  \\
\label{eq:linear.model}
\eta = g(\theta) = \ln \left (\frac{\theta}{1-\theta} \right )
& = \beta_0 x_0 + \beta_1 x_1 + \ldots
+ \beta_p x_p.
\end{align}
The focus in this section is to derive priors for the coefficients of the above logistic regression model such that the induced prior for $\theta$ is a Uniform(0,1) distribution.  These simple results will be built on in later sections to yield a Uniform(0,1) prior for $\theta$ when there are $p \ge 2$ covariates in the regression model and then are extended to the case where the induced prior is a Beta($\alpha, \beta$) for arbitrary shape parameters.  

\subsection{From $\pi(\beta)$ to $\pi(\theta)$}
Regarding the induced priors for $\theta$, for maximum simplicity
we begin with a linear model
for $\eta$ that is a constant, i.e., $\eta = \beta$:
\begin{align*}
    \eta = g(\theta) = \ln \left (\frac{\theta}{1-\theta} \right ) &= \beta.
\end{align*}
Let $\pi(\beta)$ be the prior distribution for
$\beta$, which in this simple case is also the prior for $\eta$, 
i.e., $\pi(\eta)=\pi(\beta)$. 
The induced prior for $\theta$ can be found by the
change of variable theorem \citep{2024_Siegrist}:
\begin{align*}
    \pi_\theta(\theta) & = 
    \pi_\beta (h^{-1}(\theta))
    \left | \frac{dh^{-1}(\theta)}{d\theta} \right |,
\end{align*}
where $h(\beta)=\theta$.
We define,
\begin{align*}
  \theta= h(\beta) & = \frac{\exp(\beta)}{1+\exp(\beta)}.
\end{align*}
% The inverse function is,
% \begin{align*}
%   \beta=  h^{-1}(\theta) & = 
%   \ln \left ( \frac{\theta}{1-\theta} \right ),
% \end{align*}
% which is simply the link function, $g(\theta)$, above.
% Thus,
% \begin{align*}
% \frac{d h^{-1}(\theta)}{d \theta} &= \frac{1}{\theta (1-\theta)},
% \end{align*}
which leads to the induced pdf for $\theta$:
\begin{align*}
\pi_\theta (\theta) & = \pi_{\beta} \left (\ln \left [ \frac{\theta}{1-\theta} \right ] \right )
\left | \frac{1}{\theta (1-\theta)} \right |.
\end{align*}

\paragraph{Example.} A Normal(0, $\sigma^2$) prior is
chosen for $\beta$.  The induced prior for $\theta$
is the following.
\begin{align*}
\pi(\theta) &= \frac{1}{\sqrt{2\pi \sigma^2}}
 \exp \left [ - \frac{ \left (\ln(\theta/(1-\theta)) \right )^2}{2\sigma^2} \right ] 
\frac{1}{\theta (1-\theta)} .
\end{align*}
 
\begin{figure}[h]
    \centering
\begin{subfigure}{0.48\textwidth}
    \includegraphics[width=\textwidth]{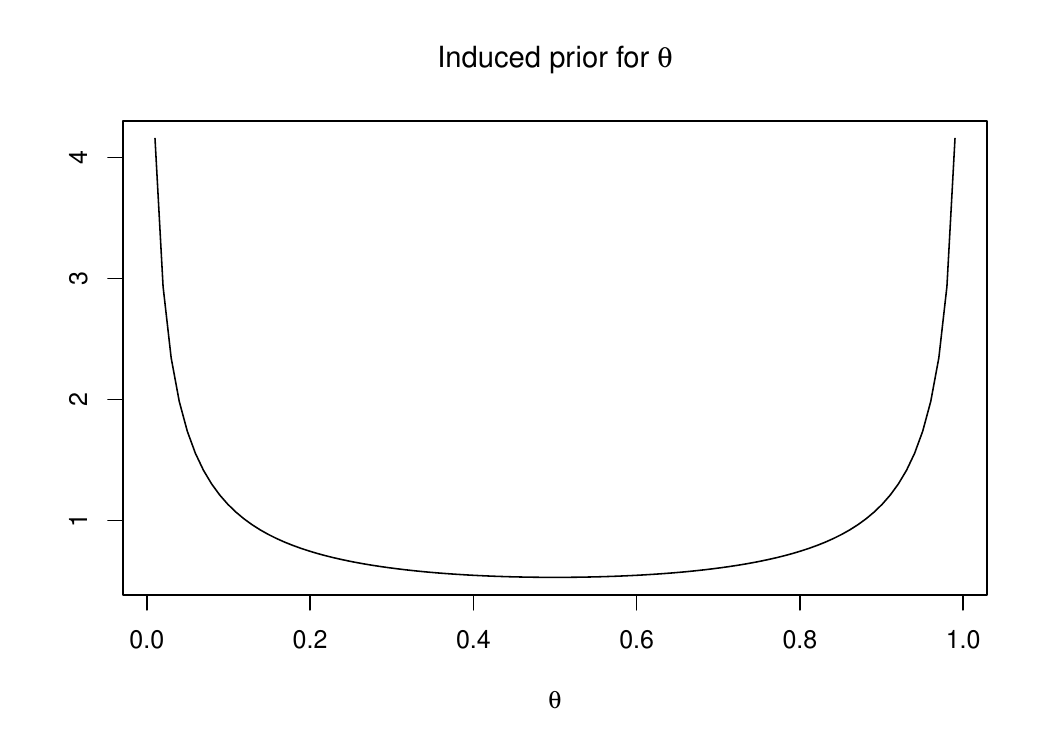}
    \caption{$\sigma^2$=3$^2$.}
     \label{F:one.induced.theta.prior}
   \end{subfigure} 
    \hfill 
   \begin{subfigure}{0.48\textwidth}
    \includegraphics[width=\textwidth]{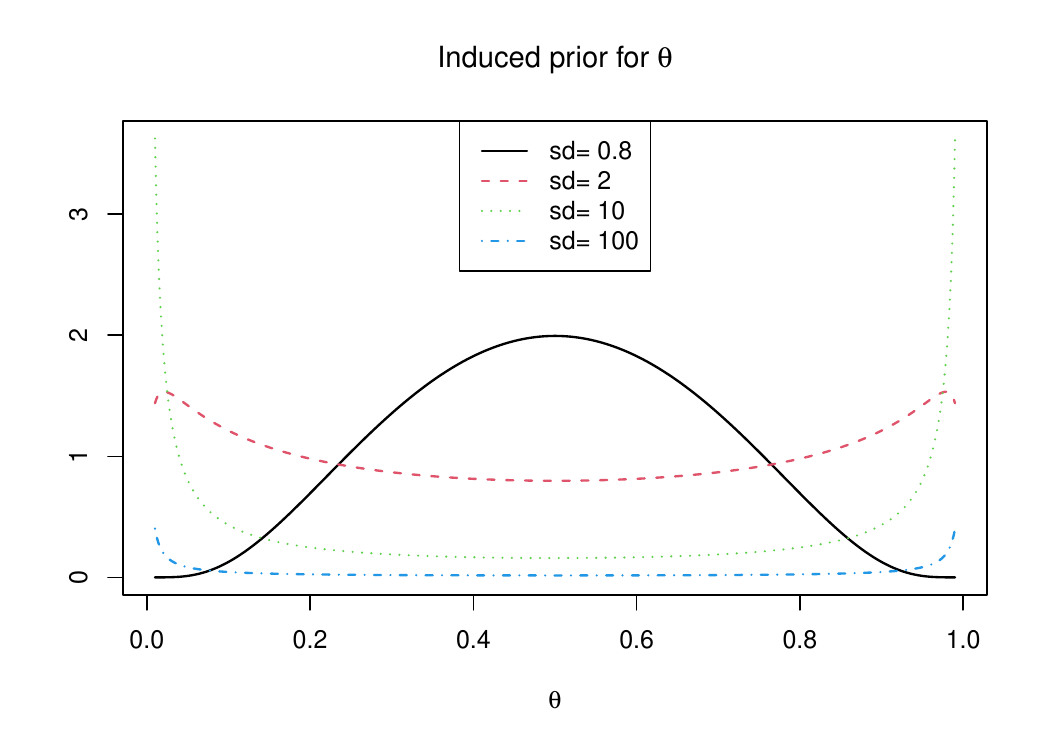}
    \caption{Increasing $\sigma^2$.}
    \label{F:several.induced.priors}  
   \end{subfigure}
    \caption{Examples of induced prior for Bernoulli($\theta$)
with logit linear model = $\beta$, where $\pi(\beta)$
is Normal($0,\sigma^2$).}
    \label{F:induced.theta.prior}
\end{figure}

Figure \ref{F:one.induced.theta.prior} is a plot of
the induced prior for $\theta$ given a 
prior of Normal$(0,\sigma^2=3^2)$ for $\beta$. This
is a ``bathtub prior'' for  
$\theta$ which concentrates the prior probability around the
extremes 0 and 1, where the concentration becomes more extreme
the larger the variance for $\beta$,
as can be seen in Figure \ref{F:several.induced.priors}.
 \cite{2012_Seaman_etal}, \cite{2018_Northrup_Gerber}
and \cite{2020_Lele} provide additional examples.

\subsection{From $\pi(\theta)$ to $\pi(\beta)$}
\begin{figure}[h]
    \centering
    \includegraphics[width=0.6\columnwidth]{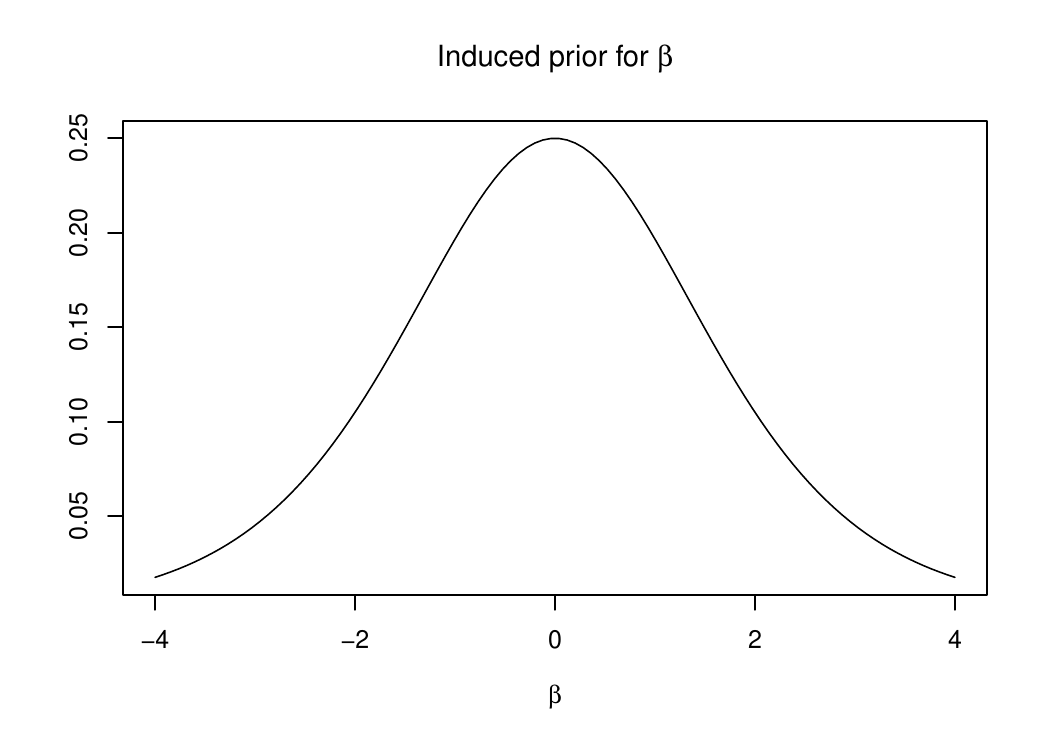}
    \caption{Induced prior for $\beta$
when $\beta$=logit($\theta$) and $\theta$
$\sim$ Uniform(0,1).}
    \label{F:induced.beta.prior}
\end{figure}

Our  objective is to determine
what the induced prior for $\beta$ is given a desired
or target prior for $\theta$.  The above process
is reversed.  For example, if the desired prior for $\theta$
is Uniform$(0,1)$, then the problem is to determine
what prior for $\beta$ would yield such a prior for
$\theta$. To begin
\begin{align*}
    \beta = g(\theta)  &= 
    \ln \left ( \frac{\theta}{1-\theta} \right ). 
\end{align*}
Then the inverse operation, $g^{-1}(\theta)$, is
$\theta = \exp(\beta)/(1+\exp(\beta))$ and 
 $\frac{dg^{-1}(\beta)}{d\beta}$  =
  $\frac{\exp(\beta)}{(1+\exp(\beta))^2}$.
Consequently the induced pdf for $\beta$ is:
\begin{align}
  \label{eq:logit.pdf}
  \pi_\beta(\beta) & = \frac{\exp(\beta)}{(1+\exp(\beta))^2},
 ~~ -\infty < \beta < \infty . 
\end{align}
Figure \ref{F:induced.beta.prior} is a plot of
the induced prior for $\beta$ given a Uniform(0,1)
prior for $\theta$.

\subsection{Logistic Distribution}
The pdf for the induced prior for $\beta$ in
Equation (\ref{eq:logit.pdf}) is a special
case of the logistic distribution. The logistic
distribution has location parameter $\mu$ and
scale parameter $s$:
\begin{align*}
\mbox{Logistic}(x; \mu, s) & = \frac{\exp \left ( -\frac{(x-\mu)}{s} \right )}
{s \left ( 1+\exp \left ( -\frac{(x-\mu)}{s}\right )
\right )^2}.
\end{align*}
The corresponding expectation and variance are as follows:
\begin{align*}
  E[X]  = \mu,    ~&~ V[X] = \frac{s^2 \pi^2}{3} . 
\end{align*}

The pdf in Equation (\ref{eq:logit.pdf}) is then
Logistic($\mu=0$, $s=1$), where due to symmetry
in the pdf around 0,
% \footnote{\textcolor{blue}{Footnotes in an article are uncommon, so may want to move this to text, or just state the symmetry result.  I spelled it out for my own benefit as it wasn't immediately obvious to me.} Showing the symmetry
% for the special case of $\mu=0$ and $s=1$, the ratio of
% of the pdfs evaluated at $x$ and -$x$ is
% \begin{align*}
%     \frac{\exp(-x)/(1+\exp(-x))^2}
%     {\exp(-(-x))/(1+\exp(-(-x)))^2} &=
%     \frac{\exp(-x)(1+2\exp(x)+\exp(2x))}
%     {\exp(x)(1+2\exp(-x)+\exp(-2x)} =
%     \frac{\exp(-x)+2+\exp(x)}{\exp(x)+2+\exp(-x)}=1
% \end{align*}.}
the pdf is the same for $\beta$
and $-\beta$.  $E[\beta]=0$ (as Figure \ref{F:induced.beta.prior} indicates) and $V[\beta]$=$\frac{\pi^2}{3}$.
We can conclude from this that if a Logistic$(0,1)$ prior is used for $\beta$, the induced
 prior for $\theta$ is Uniform$(0,1)$.
  Depending on the context, the target induced prior for $\theta$ may not be Uniform$(0,1)$, e.g., it might be Beta($\alpha, \beta$) and this is later addressed.

\subsection{Normal approximation to Logistic(0,1)}\label{sec:normal-logistic}
\begin{figure}[h]
    \centering
    \includegraphics[width=0.6\columnwidth]{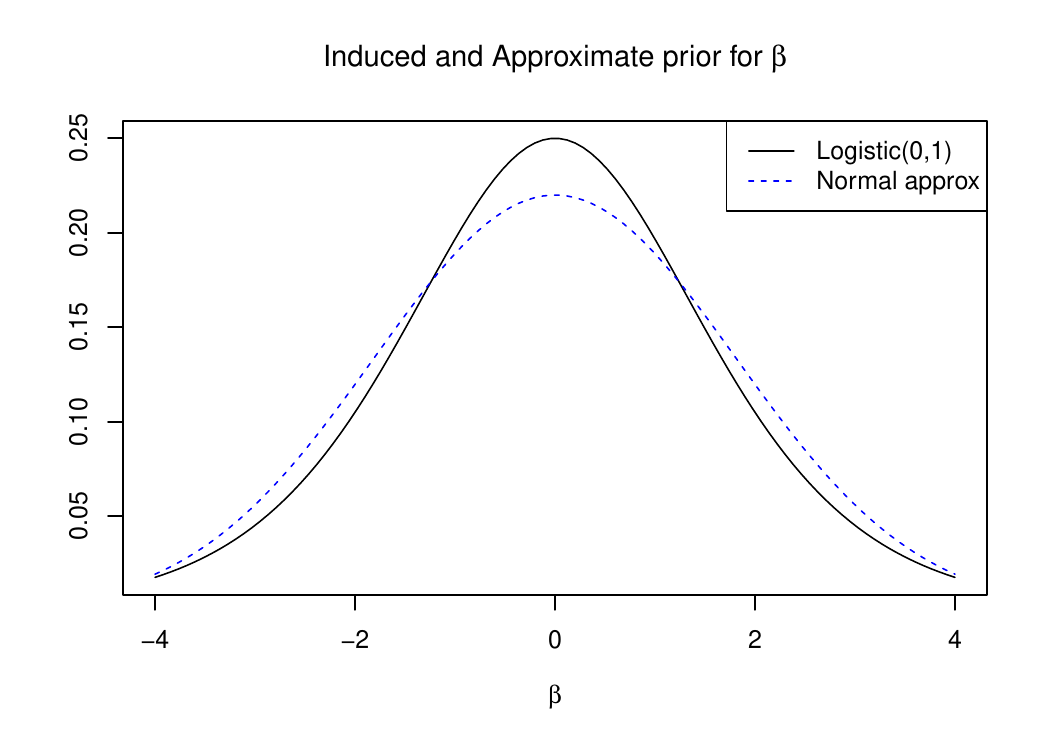}
  \caption{Induced  prior (Logistic(0,1)) for $\beta$ in logit linear model for a Bernoulli parameter $\theta$ when $\pi(\theta)$
is Uniform(0,1) and a Normal(0,$\frac{\pi^2}{3}$)
approximation.}
    \label{F:induced.approx.prior}
\end{figure}

   Given the shape of the Logistic(0,1) distribution and its symmetry about zero, a Normal(0, $\frac{\pi^2}{3}$)
distribution could be used as an approximation.
As Figure \ref{F:induced.approx.prior} shows the
tails of the normal are heavier.

%% file: 3_Mean_Variance_Prior_Logistic_Reg_Coef.tex
\section{\label{sec:mean.var.determination}Moment
matching for $\pi(\beta)$'s
for multivariate Logistic regression}
 Now we consider the more general logistic regression
 setting where $\theta$ is a function of $p$ covariates, as shown in Equation (\ref{eq:linear.model}).
 From Section \ref{sec:simple.logistic},
 if the prior for $\eta$ is Logistic$(0,1)$, then
 the induced prior for $\theta$ is Uniform$(0,1)$. 
 Therefore we need
 \begin{align*}
    \eta = \beta_0   + \beta_1 x_1 + \ldots
+ \beta_p x_p & \sim \mbox{Logistic}(0,1),
 \end{align*}
 where we have assumed $x_0=1$. 
 There are two complications here. Firstly, 
 given $(p+1)$ $\beta$ parameters, there are $p+1$ prior
 distributions. One is that there is a many-to-one mapping from the $\beta$s to the single $\theta$. In 
 theory there are an infinite number of combinations of prior distributions for the $\beta$'s that
 induce a Logistic$(0,1)$ distribution.  Secondly, the values of $x_0, x_1, \ldots, x_p$
 will affect the distribution of $\eta$. 
 Here we will examine approximate solutions to these two problems where we aim to match the expected
 value and variance for $\eta$.    
 
 \subsection{Linear model with all $x_i$=1}\label{sec:x1}
 We begin with an unrealistic setting to make clearer the approach taken by assuming that the linear model for $\eta$ is simply a sum of $p+1$ coefficients, equivalently assume that all $x_i$=1:
 \begin{align*}
 %\label{eq:sum.betas}
    \eta = g(\theta) &= \beta_0 + \beta_1 \times 1 + \ldots 
    + \beta_p \times 1 = \sum_{i=0}^p \beta_i.
 \end{align*}
 Hence there is a $p+1$ dimensional prior,
  $\pi(\beta_0, \ldots, \beta_p)$. The
 many-to-one problem can be bypassed by assuming that
 the prior distributions are all the same and
 independent.  By
 moment matching we can determine what the mean,
 $\mu_\beta$,  and variance, $\sigma_{\beta}^2$, of the
 in-common prior, denoted $\pi(\beta)$ distribution,
 should  be.
 \begin{align*}
E[\eta] = 0 &= E[\beta_0] + E[\beta_1] + \ldots 
E[\beta_p] = (p+1) \mu_\beta \\
V[\eta] = \frac{\pi^2}{3} & = V[\beta_0] + V[\beta_1]
\ldots + V[\beta_p] = (p+1) \sigma_{\beta}^2
 \end{align*}
 So that $\mu_\beta=0$ and $\sigma^2_{\beta}$ = 
 $\frac{\pi^2}{3(p+1)}$. When $p = 0$, we obtain the variance of $\frac{\pi^2}{3}$ considered in Section \ref{sec:normal-logistic}.
 
 % Later we propose an alternative solution to the many-to-one problem
 % by first specifying a prior for the intercept,
 % as that reflects average responses, $\pi_{\beta_0}(\beta_0)$.  Then assume   the
 % remaining priors, $\pi_{\beta_i}(\beta_i)$, $i$=1,$\ldots,p$, are
 % again iid and solve for them given $\pi(\eta)$ and
 % $\pi_{\beta_0}(\beta_0)$. 
 
 \subsection{Moment matching with standardised covariates}
 We now examine the more realistic case where there
 are covariates in the linear model.  We will, however,
 assume that the
 covariates, $x_1$, $\ldots$, $x_p$, are standardised,
$x_i$= $(x_i^*-\bar{x_i^*})/s_{x_i^*}$, 
where $x_i^*$ are the
original values, thus $\bar{x_i}$=0 and  $s_{x_i}$= 1.
To bypass the many-to-one problem,
we  assume ``some'' in-common, and
independent  prior distributions. To begin we will
assume that all the coefficients $\beta_0$, $\ldots$,
$\beta_p$ are iid with prior $\pi_{\beta}(\beta)$.
   
 We again use moment matching to determine
 what conditions are necessary for $E[\eta]$=0
 and $V[\eta]$=$\pi^2/3$ to be satisfied.
 \begin{align*}
E[\eta] = 0 &= E[\beta_0   + \beta_1 x_1 + \ldots
+ \beta_p x_p] \\
%&= E \left [ E_{\bfx|\beta} 
% [\beta_0   + \beta_1 x_1 + \ldots
%+ \beta_p x_p \left |  \beta's \right .] \right ] \\
%&= E \left [ \beta_0 + \sum_{i=1}^p \beta_i E[X_i] \right ] \\
& = E[\beta_0] + \sum_{i=1}^p E[\beta_i] E[X_i] . 
 \end{align*}
 Given that the covariate values have been standardised, their
 expected values, $E[X_i]$, equal 0, so that 
 the second term above equals 0. This result arises from the expectation   being calculated over a finite space
 with at most $n$ values for each $x_i$ and assuming
 implicitly random draws of each $x_i$ independent
 of $x_j$, $i \ne j$, which does not recognize any
 within sample dependencies.  However, to ensure $E[\eta]$=0, 
$E[\beta_0]$ must also equal 0.  If the $\beta$'s
are iid, their expected values would also equal 0.
 
We next examine $V[\eta]$.  In the following we 
assume that $V[x_i]$=1.
\begin{align*}
V[\eta] = \frac{\pi^2}{3} &= V \left [ \beta_0 + 
 \sum_{i=1}^p \beta_i x_i \right ] \\
&= V_\beta E_{\bfx|\beta} \left [ \beta_0 + 
 \sum_{i=1}^p \beta_i x_i \left | \beta \right . \right ]  +
 E_\beta V_{\bfx|\beta}\left [ \beta_0 + 
 \sum_{i=1}^p \beta_i x_i \left |  \beta  \right . \right ] \\
 &= V_\beta \left [ \beta_0 + 
 \sum_{i=1}^p \beta_i E[x_i] \right ]  +
 E_\beta  \left [  
 \sum_{i=1}^p \beta_i^2 V[x_i] \right ] \\
 &= V_\beta[\beta_0] + 0 + 
 \left [ \sum_{i=1}^p E_\beta[\beta_i^2] \times 1\right ] \\
%
%&= \sigma^2_\beta + p\sigma^2_\beta \\
 & = \sigma^2_\beta(1+p) .
\end{align*}
To match the Logistic$(0,1)$ variance of $\frac{\pi^2}{3}$, we require
\begin{align*}
\sigma^2_\beta & = \frac{\pi^2}{3(p+1)}.
\end{align*} 
Therefore, as for the case with $x_i$=1, $\mu_\beta$=0 and $\sigma^2_{\beta}$ = 
 $\frac{\pi^2}{3(p+1)}$. The reason for the identical results is that the standardized variance of the $x_i$'s, (namely, 1), is the same as the assigned constant of 1 in the previous case. \citet{2003_Newman} proposed the identical normal prior distributions for an analysis of release-recovery data but the derivation details were not shown.

%% file: 4_Arbitrary_beta_theta.tex
\section{\label{sec:arbitrary.beta}Arbitrary Beta distribution
for $\theta$}
\begin{figure}[h]
    \centering
    \includegraphics[width=0.8\linewidth]{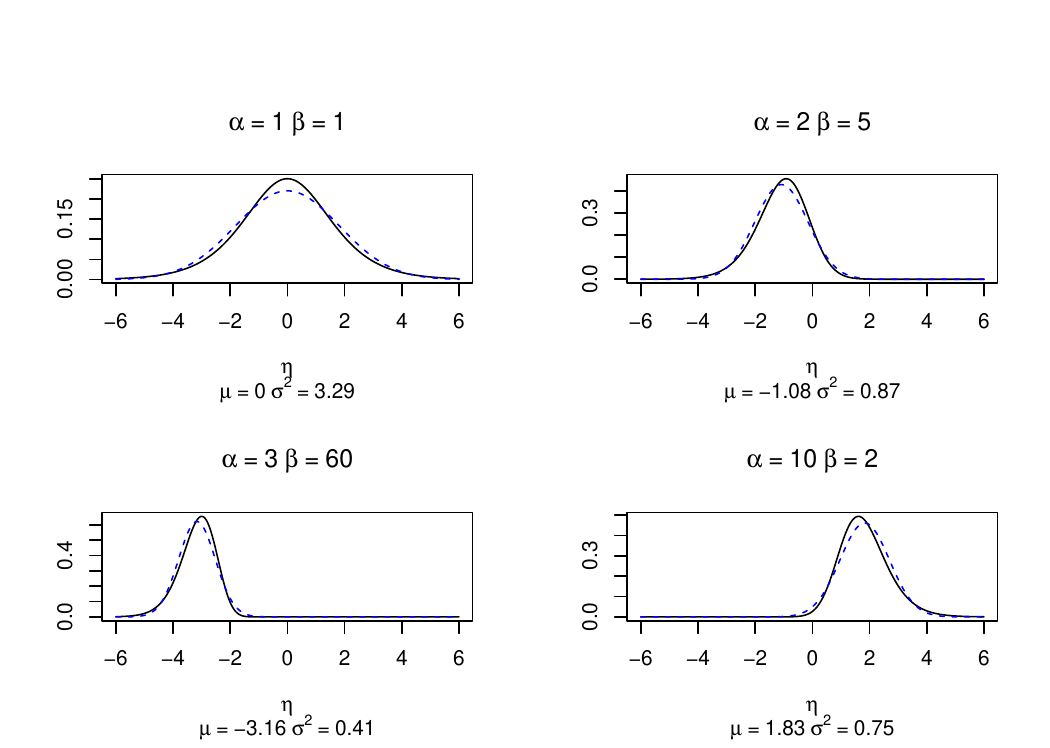}
    \caption{Density plots for $\eta=$ logit$(\theta)$ where $\theta \sim Beta(\alpha,\beta)$ for a variety of Beta distribution shape
    parameters. Numerically calculated averages and variances are shown
    below each plot and a normal approximation is shown
    with a dashed blue line.}
    \label{F:eta.pdf}
\end{figure}

Now we consider the case where an arbitrary Beta distribution has been chosen for the Bernoulli parameter $\theta$ and to determine an induced distribution for the logit transform.
If the prior for $\theta$ is Beta($\alpha, \beta$), and
$\eta$ = logit($\theta$),  the induced pdf for $\eta$
is the following.
\begin{align}
\label{eq:induced.eta.arbitrary.beta}
p(\eta) &= \frac{1}{B(\alpha,\beta)} \frac{\exp(\eta \times \alpha)}{(1+\exp(\eta))^{\alpha+\beta}},
\end{align}
where $B(\alpha,\beta)$ is the beta function: 
$\frac{\Gamma(\alpha) \Gamma(\beta)}{\Gamma(\alpha+\beta)}$,
which is the pdf of the Type IV generalized logistic regression \citep{1995_Johnson_etal}.  Examples of the pdf for
$\eta$, eq'n \ref{eq:induced.eta.arbitrary.beta},
given different
Beta($\alpha,\beta)$ distributions for $\theta$,
are shown in Figure \ref{F:eta.pdf}.

 The values of $E[\eta]$ and
$Var[\eta]$  shown in Figure \ref{F:eta.pdf} were found by numerically integrating the following expected values:
\begin{align*}
%\label{eq:E.eta.arb.beta}
E[\eta^k] &= \frac{1}{B(\alpha,\beta)} 
 \int_{-\infty}^{\infty} \eta^k \frac{\exp(\eta \times \alpha)}{(1+\exp(\eta))^{\alpha+\beta}} d\eta,  ~~ k=1,2.
\end{align*}
Hereafter we denote $E[\eta]$ by $\mu_\eta$ and
$Var[\eta]$ by $\sigma^2_\eta$. 

\subsection{Normal approximation intercept only model}
As for the special case of Uniform$(0,1)$, i.e.
Beta$(1,1)$, a normal approximation to the distribution 
for $\eta$ is to moment match $\mu_\eta$ and
$\sigma^2_\eta$. 

For an intercept
only model, $\eta = \beta_0$, and the  
prior for $\beta_0$ is
\begin{align*}
\beta_0 & \sim \mbox{Normal} \left ( \mu_\eta, \sigma_\eta \right ).
\end{align*}
Examples of such normal approximations to the pdf for $\eta$ are
shown by dashed blue lines in Figure \ref{F:eta.pdf}, which
show the adequacy
of the approximation is a function of $\alpha$ and $\beta$.

\subsection{Normal approximation: Multiple covariates}
In the case of one or more covariates, as for the Uniform$(0,1)$ case, there is again the many-to-one problem where there are an infinite
number of combinations of normal priors where the expected
value and variance for the sum of the priors would equal
$\mu_\eta$ and $\sigma^2_\eta$. Furthermore, given that standardizing the covariates to have mean zero implies $E[\beta_i X_i]=0$ 
the mean $\mu_\eta$ is solely found in the prior mean for 
the intercept $\beta_0$.  Again, we can avoid the many-to-one
problem by using the
$p$ independent and identically distributed normal distributions  for the slope coefficients.  Namely, 
\begin{align}
\label{eq:b0.prior.arbitrary.theta.prior}
 \beta_0 & \sim \mbox{Normal} \left ( \mu_\eta, 
 \frac{\sigma^2_\eta}{p+1}\right ); \\
 \label{eq:bi.prior.arbitrary.theta.prior}
 \beta_i & \sim  \mbox{Normal} 
 \left ( 0, \frac{\sigma^2_\eta}{p+1} \right ), \qquad i=1,\ldots,p.
\end{align}
Setting the mean of the priors for the $\beta_i$ terms equal to 0 is arbitrary and can be any value since $E[\beta_i X_i] = 0$. 
Then
\begin{align*}
E[\eta] & = E[\beta_0] + \sum_{i=1}^p E[\beta_i x_i] =
  \mu_\eta  +
 \sum_{i=1}^p 0 = \mu_\eta; \\
 Var[\eta] & = Var[\beta_0] + \sum_{i=1}^p Var[\beta_i x_i]
 = \frac{\sigma^2_\eta}{p+1} + 
 \sum_{i=1}^p \frac{\sigma^2_\eta}{p+1} = \sigma^2_\eta.
\end{align*}
As $p$ increases, the variances for both the intercept and the
slope coefficients are equally increasingly concentrated 
around, or \textit{shrinking} toward, $\mu_\eta$ or 0. We next present an alternative which does not cause such shrinkage in the prior for the intercept.

We note that while there are different classes of so-called shrinkage priors , e.g., spike and slab priors, the intent differs somewhat, to shrink small effects to zero while maintaining true large effects \citep{2019_VanErp_etal}, while the intent is here is to yield a desired induced distribution on a particular parameter.

\subsection{\label{sec:Handling.Intercept}Distinct handling
of the intercept}
In some cases there may be a prior opinion regarding the 
value of
$\theta$ even if the covariates have no effect, or for an ``average'' covariate setting, and a
distinct handling of the intercept can be considered. In
particular one may not want to have the number of covariates
affecting the prior for the intercept, i.e., having the prior for $\beta_0$ become increasingly concentrated around zero. 

We begin by assuming a more general prior for $\theta$, Beta($\alpha,\beta$), rather than the special case dealt with so far, namely $\alpha = \beta = 1$ or Uniform$(0,1)$.  Direct specification of the $\alpha$ and $\beta$ parameters could be done as well but thinking in terms of the average value for $\theta$ and its variation may be easier. Specifically, we first specify the expected value of $\theta$ and a measure of its variation, the  coefficient of variation (CV), and then determine what values of $\alpha$ and $\beta$ match these specified values.  
For example, if we specify $E[\theta] = 0.7$ and 
$CV[\theta]=0.30$, then the prior for $\theta$ is Beta($2.633, 1.129$).   Given the  Beta distribution parameters, 
and referring to Equation (\ref{eq:induced.eta.arbitrary.beta}), 
the expected value and
variance of $\eta$ can be numerically calculated: in this
example $\mu_\eta$ $\approx$ 1.146 and $\sigma^2_\eta$
$\approx$ 1.785.  Again assuming that the priors for $\beta_0$ and $\beta_i$, $i=1,\ldots,p$ are Normal distributions, the resulting $E[\eta]$ and $Var[\eta]$ guide the choice of the Normal distribution parameters. 

We note that with the exception of the case where $\alpha$=$\beta$, thus $E[\theta]$=0.5, and then $E[\eta]$ = 0, the contribution of $\mu_\eta$ to the linear combination (Equation \ref{eq:linear.model}) must all come from the mean for the prior for the intercept, $\beta_0$. This is because the covariates have been standardized to have mean 0, then $E[\beta_i x_i]$ = 0 no matter what the value of the expected value for $\beta_i$ in the normal distribution. 

To bypass the many-to-one issue we again assume that the priors for the slope coefficients are identical.  Let  $\mu_0$ and $\sigma^2_0$ denote the mean and variance for the normal prior for $\beta_0$. For the normal prior for $\beta_i$, $\mu_1$ denotes the normal mean for the $\beta_i$'s, which can be anything given it does not affect the expected value of the linear combination, of $\eta$.  Here for simplicity we will set $\mu_1$=0, and $\sigma^2_1$ denotes the variance for prior for $\beta_1$. While there might be some argument for $\mu_1 \ne 0$, we cannot think of one.  Subsequently,
\begin{align}
\label{eq:b0.prior.arbitrary.theta.prior.wtd}
\beta_0 & \sim \mbox{Normal} \left ( \mu_\eta, \sigma^2_0 \right ), \\
\label{eq:bi.prior.arbitrary.theta.prior.wtd}
\beta_1 & \sim \mbox{Normal} \left ( 0, \sigma^2_1 \right ).
\end{align}
The constraint for the linear combination is that its expected value equals $\mu_\eta$ and its variance equals
$\sigma^2_\eta$. Redisplaying the linear combination for convenience, with a randomly drawn sequence of values for
$x_{i,j}$
\begin{align*}
\eta_j & = \beta_0 + \sum_{i=1}^p \beta_i x_{i,j} ,
\end{align*}
with $E[\eta_j]$ = $\mu_\eta$ and 
\begin{align*}
Var[\eta_j] & = \sigma^2_0 + p \sigma^2_1,
\end{align*}
where the latter sum must equal $\sigma^2_\eta$. 
Given values for $\mu_\eta$ and $\sigma^2_\eta$,
once   $\sigma^2_0$ are specified then the variance
for the slope coefficient priors is 
\begin{align*}
%\label{eq:prior.var.beta.i}
    \sigma^2_1  &= \frac{\sigma^2_\eta-\sigma^2_0}{p}.
\end{align*}
Given the assumption of identical priors for all slope 
coefficients the only decision is to specify  
$\sigma^2_0$. There are numerous ways this can 
be done and that shown below is just one possibility.

 If the aim is that the number of covariates does not
affect the prior for $\beta_0$, in particular the prior variance for $\beta_0$, then $\sigma^2_0$
should not be a function of $p$.  One approach is to 
define $\sigma^2_0$ relative to $\sigma^2_\eta$.
For example specify a constant, $k$, $0 < k < 1$, 
and set $\sigma^2_0$ = $k \sigma^2_\eta$. Then
\begin{align*}
\sigma^2_1 & = \frac{(1-k) \sigma^2_\eta}{p}.
\end{align*}
The choice of $k$ is discussed in Section \ref{sec:Discussion}.

%% file: 5_Simulations_0.tex
\section{\label{sec:simulations}Simulations}
In this section, we present the results of three simulation experiments in as many scenarios (conducted using \texttt{R} \citep{2024_Rlang}). We consider:
\begin{enumerate}
\item Scenario 1: one covariate and a Uniform(0,1) prior for $\theta$.
\item Scenario 2: three covariates and a Uniform(0,1) prior for $\theta$
\item Scenario 3:  three covariates
and a Beta($\alpha,\beta$) prior for $\theta$, where $\alpha$, $\beta$ $\ne$ 1.
\end{enumerate}
 
In all three scenarios different types of Normal priors were used for the intercept $\beta_0$ and the slope coefficient(s).  We refer to the ``Vague prior'' such that there are identical diffuse Normal(0, $\sigma^2$) priors used for the intercept and slope coefficients where $\sigma^2=1000^2$; and the ``Logistic prior'', where $\sigma$ is based on the induced logistic distribution,  the details of which vary depending on the number of covariates, and the Beta($\alpha, \beta)$ prior for $\theta$.  For example, for the Logistic prior, when the induced prior for $\theta$ is Uniform(0,1), and there are $p$ covariates, then
$\sigma$ = $\sqrt{\frac{\pi^2}{3(1+p)}}$. Finally, when the prior for the intercept is handled differently (Section \ref{sec:Handling.Intercept}), the prior is called the ``Weighted prior''.

For each scenario, we performed experiments for single samples (detailed in the Appendix, as well as on multiple datasets to allow frequentist analyses of the performance.

% \newpage
\input{5_Simulations_1}

% \newpage
\input{5_Simulations_2}

\input{5_Simulations_3}

%% file: 5_Simulations_1.tex
%----------------------------------------------------
\subsection{\label{sec:scenario1}Scenario 1: One covariate and $\theta \sim U(0,1)$}

%\subsubsection{\label{sec:scenario1.single}Single dataset}
Multiple data sets of Bernoulli responses were generated from the following model:
\begin{equation}
 \begin{aligned}
    y_i & \sim \mbox{Bernoulli}(\theta_i), ~~ i=1,\ldots, 15 \\
logit(\theta_i) & =   -0.5 + 0.3 x_i, ~~ i=1,\ldots,15,
 \end{aligned}
\label{eq:model_scenario_1}
\end{equation}
where the values of $x$ were simulated from a 
Gamma(3,0.2), and then standardised. In the Appendix, the resulting response and covariate values are shown in Table \ref{T:univariate.data} and the functional relationship between $\theta$ and $x$ is shown in Figure \ref{F:univariate.theta}.

We compared the following two priors for $\beta_0$ and $\beta_1$.
\begin{equation}
\begin{aligned}
\mbox{Vague:    } \beta_0, \beta_1 & \sim \mbox{Normal}(0,1000^2); \\
\mbox{Logistic: }\beta_0, \beta_1 & \sim 
\mbox{Normal} \left (0, \frac{\pi^2}{3 \times 2} = 1.283^2 \right ).
\end{aligned}
\label{eq:prior_scenario_1}
\end{equation}
The induced priors for $\theta$ are shown in Figure \ref{F:induced.theta.priors.univariate} in the Appendix.

We simulated 100 datasets from the model with $n=15$, $n=50$ and $n=100$, and computed the following indexes: the mean squared error from the maximum likelihood estimate ($\text{MSE}^*$), the mean squared error from the posterior mean (MSE), and the coverage of the 95\% credible interval. The above mean square errors are based on the following formula:
$$\text{MSE}(\hat{\theta})=\frac{1}{n}\sum_{i=1}^n(\hat{\theta}_i-\theta)^2,$$
where $\hat{\theta}$ is the point estimate (i.e. the posterior mean or MAP, for example) and $\theta$ is either the MLE (for the $\text{MSE}^*$) or the true value of $\theta$ (for the MSE).

%---- n=15 -------
For the case $n$=15, from Table \ref{T:FrequentistAnalysis_1A} we note that the MSE for the Logistic prior is smaller (about one order of magnitude) than the one computed for the Vague normal prior. In addition, while the posterior coverage for the proposed prior is within the expected limits, under the Vague   prior we note a slightly lower than expected coverage for the intercept (0.92) and considerably lower coverage for the slope (0.78).

\begin{table}[h!]
\centering
\begin{tabular}{cccccccc}
\hline
          & & \multicolumn{3}{c}{Vague}       & \multicolumn{3}{c}{Logistic}  \\
Parameter & Truth & $\text{MSE}^*$ & MSE & Cov  & $\text{MSE}^*$ & MSE & Cov  \\
\hline
$\beta_0$        & -0.50 & 0.0253    & 1.240      & 0.92 & 0.0123    & 0.2371     & 0.97 \\
$\beta_1$        & 0.30 & 0.0347    & 1.094      & 0.78 & 0.0203    & 0.2005     & 0.93 \\
\hline
\end{tabular}
\caption{$\text{MSE}^*$, MSE and coverage of the 95\% posterior credible interval (Cov), for $n=15$, for the case with one covariate.}
\label{T:FrequentistAnalysis_1A}
\end{table}

An interesting comparison is between the average MLEs and average posterior means for both parameters. The MLEs are -0.6184 and 0.4203, for $\beta_0$ and $\beta_1$, respectively. The average posterior mean for the intercept is -0.7775 and -0.5076, for the Vague prior and the Logistic prior, respectively, and for the  slope coefficient, we have 0.6065 and 0.2780 for the Vague and the Logistic, respectively. We note that the distance between the MLEs and the posterior means are smaller for the Logistic prior than for the Vague prior.

%---- n=50 -------
For the larger sample size $n=50$, the corresponding results are presented in Table \ref{T:FrequentistAnalysis_1B} where, as expected due to increasing data having more influence on the posterior distributions, the MSE from the MLE and from the posterior mean are lower.   Note that, unlike in the case for $n=15$, the coverage of the coefficients under the Vague prior are close to the nominal value of 95\%.

\begin{table}[h!]
\centering
\begin{tabular}{cccccccc}
\hline
          & & \multicolumn{3}{c}{Vague}       & \multicolumn{3}{c}{Logistic}  \\
Parameter & Truth & $\text{MSE}^*$ & MSE & Cov  & $\text{MSE}^*$ & MSE & Cov  \\
\hline
$\beta_0$  & -0.50      & 0.0007    & 0.1165     & 0.93 & 5.00E-04  & 0.0989     & 0.94 \\
$\beta_1$  & 0.30      & 0.0017    & 0.1189     & 0.94 & 2.00E-04  & 0.0972     & 0.96 \\
\hline
\end{tabular}
\caption{$\text{MSE}^*$, MSE and coverage of the 95\% posterior credible interval (Cov), for $n=50$, for the case with one covariate.}
\label{T:FrequentistAnalysis_1B}
\end{table}
In this case, the average MLE for the intercept is -0.5104 and  for the  slope coefficient is 0.3184. The average posterior means are, for the intercept, -0.5290 (Vague prior) and -0.4950 (Logistic prior) and, for the  slope coefficient, 0.3425 (Vague prior) 0.3165 (Logistic prior). Here, while the distances of the MLE and the posterior means for the intercept are virtually the same, the ones for the  slope coefficient are smaller for the Logistic prior.

%---- n=100 -------
To conclude this experiment, we have run the above simulation with a larger sample size ($n=100$). The results are presented in Table \ref{T:FrequentistAnalysis_1C}, where we note that the priors' behaviours are virtually the same as the information in the data dominates. Again, the overall performance of the Logistic prior is better than the Vague prior, in terms of distance between the MLE and the posterior mean, the difference diminishes, as one would expect as the sample size increases.

\begin{table}[h!]
\centering
\begin{tabular}{cccccccc}
\hline
          & & \multicolumn{3}{c}{Vague}       & \multicolumn{3}{c}{Logistic}  \\
Parameter & Truth & $\text{MSE}^*$ & MSE & Cov  & $\text{MSE}^*$ & MSE & Cov  \\
\hline
$\beta_0$  & -0.50      & 1.00E-04  & 0.0496     & 0.94 & 1.00E-04  & 0.0458     & 0.95 \\
$\beta_1$  & 0.30       & 3.00E-04  & 0.0497     & 0.97 & 0.00E+00  & 0.0457     & 0.96 \\
\hline
\end{tabular}
\caption{$\text{MSE}^*$, MSE and coverage of the 95\% posterior credible interval (Cov), for $n=100$, for the case with one covariate.}
\label{T:FrequentistAnalysis_1C}
\end{table}
For this final simulation study, the average MLE for the intercept is -0.5265 and for the  slope coefficient is 0.2914. The average posterior means are, for the intercept, -0.5355 (Vague prior) and -0.5206 (Logistic prior) and, for the  slope coefficient, 0.3013 (Vague prior) and  0.2914 (Logistic prior).

%% file: 5_Simulations_2.tex
%----------------------------------------------------
\subsection{\label{sec:scenario2}Scenarios 2 and 3: Three covariates} 

In the second  and third scenarios, we drew samples from the below model given in Equation \eqref{eq:multiple.covariates}, considering $\theta\sim U(0,1)$ (Scenario 2), and its generalisation $\theta\sim \text{Beta}(\alpha,\beta)$ with distinct handling of the intercept (Scenario 3):
\begin{align}
\label{eq:multiple.covariates}
logit(\theta) & = 1.1 + 0.3 x_1 -0.6 x_2 +0.02 x_3,
\end{align}
where the covariates were simulated independently from three Gamma distributions, Gamma(10,2), Gamma(12,6), and Gamma(3,3), with the resulting $\theta$ values ranged from 0.39 to 0.93.

The prior distributions we compared are the Vague prior and the Logistic prior (see equations \eqref{eq:b0.prior.arbitrary.theta.prior} and \eqref{eq:bi.prior.arbitrary.theta.prior}) for the three covariates setting in Scenario 2.
\begin{equation}
\begin{aligned}
\mbox{Vague:    } \beta_0, \beta_1, \beta_2, \beta_3 & \sim \mbox{Normal}(0,1000^2) \\
\mbox{Logistic: }\beta_0, \beta_1, \beta_2, \beta_3 & \sim 
\mbox{Normal} \left (0, \frac{\pi^2}{3 \times (1+3)} = 0.907^2 \right ),
\end{aligned}\label{eq:prior_scenario_2}
\end{equation}

and, for Scenario 3, the additional prior
\begin{equation}
\begin{aligned}
    \beta_0&\sim \text{Normal}(1.150,0.4\times 1.843=0.7372)\\
    \beta_i&\sim \text{Normal}\left(0,\frac{0.6\times 1.843}{3}=0.3686\right),\qquad i=1,2,3.
\end{aligned}\label{eq:weightedprior}
\end{equation}

In the Appendix, it is possible to find the induced $\theta$ priors, as well as the illustration of the model applied to a single sample.\\

The frequentist analysis for Scenario 2 was performed on 100 samples of sizes $n=30$, $n=50$ and $n=100$, respectively.  We defer the discussion of the results below where
the Vague, Logistic, and Weighted priors are shown together.

%% file: 5_Simulations_3.tex
Similrly to Scenario 2, the frequentist analysis of Scenario 3 was performed on 100 samples of sizes $n=30$, $n=50$ and $n=100$, respectively.

%---- n=50 -------
Table \ref{T:Freq_3cov_n50}, for $n$=50,  shows the $\text{MSE}^*$, the MSE and interval coverage for the three priors. It appears that both the Logistic prior and Weighted prior perform better than the Vague prior; this is obvious by noticing that the MSE is systematically smaller for the two proposed priors for all the parameters. If we compare the coverage, we notice that the Weighted prior tends to have more extreme results, as it is either close to 100\% or 90\%, suggesting a less precise posterior. Given that the Weighted prior assumes different prior for the intercept and the slope coefficients, an increase in uncertainty is expected.

\begin{table}[h!]
\centering
\begin{tabular}{cccccccccc}
\hline
          & \multicolumn{3}{c}{Vague}       & \multicolumn{3}{c}{Logistic}  & \multicolumn{3}{c}{Weighted}      \\
Parameter & $\text{MSE}^*$ & MSE & Cov  & $\text{MSE}^*$ & MSE & Cov  & $\text{MSE}^*$ & MSE & Cov  \\
\hline
$\beta_0$        & 0.18    & 1.30     & 0.89 & 0.34    & 0.11     & 0.96 & 0.36     & 0.09     & 0.97 \\
$\beta_1$        & 0.03    & 0.45     & 0.89 & 0.07    & 0.10     & 0.97 & 0.20     & 0.05     & 0.97 \\
$\beta_2$        & 0.04    & 0.45     & 0.90 & 0.10    & 0.10     & 0.96 & 0.28     & 0.12     & 0.89 \\
$\beta_3$        & 0.10    & 1.05     & 0.91 & 0.28    & 0.09     & 0.99 & 0.38     & 0.03     & 1.00\\
\hline
\end{tabular}
\caption{$\text{MSE}^*$, MSE and coverage of the posterior 95\% credible interval for the Vague prior, the Logistic prior and the Weighted prior. The summaries refer to 100 samples of size $n=50$.}
\label{T:Freq_3cov_n50}
\end{table}

For the above simulation study, the average MLEs, and the average posterior means (under each prior) are reported in Table \ref{T:Freq_3cov_n50_summaries}.

\begin{table}[h!]
\centering
\begin{tabular}{cccccc}
\hline
          &      &       & \multicolumn{3}{c}{Posterior Means} \\
Parameter & TRUE & MLE   & Vague      & Logistic     & Wted      \\
\hline
$\beta_0$ & 1.5  & 1.6905  & 1.9587     & 1.3887         & 1.4235      \\
$\beta_1$ & 0.3  & 0.4563  & 0.5580     & 0.3324         & 0.1793      \\
$\beta_2$ & -0.6 & -0.6801 & 0.7931     & -0.5125        & -0.2997     \\
$\beta_3$ & 0.02 & 0.0441  & 0.1010     & 0.0115         & 0.0013   \\
\hline
\end{tabular}
\caption{Average maximum likelihood estimates, and average posterior means for the $n=50$ simulation study.}
\label{T:Freq_3cov_n50_summaries}
\end{table}

%---- n=30 -------
Similarly to the case with one covariate, we examined (and compared) the prior performances for a relatively small sample size. Using the rule of thumb of having ten data points per covariate, we have drawn 100 samples of size $n=30$. The results, in terms of $\text{MSE}^*$, MSE and posterior coverage, are reported in Table \ref{T:Freq_3cov_n30}.

\begin{table}[h!]
\centering
\begin{tabular}{cccccccccc}
\hline
          & \multicolumn{3}{c}{Vague}       & \multicolumn{3}{c}{Logistic}  & \multicolumn{3}{c}{Weighted}      \\
Parameter & $\text{MSE}^*$ & MSE & Cov  & $\text{MSE}^*$ & MSE & Cov  & $\text{MSE}^*$ & MSE & Cov  \\
\hline
$\beta_0$ & 0.8154    & 3.0434     & 0.75 & 0.5970    & 0.1512     & 0.96 & 0.5543    & 0.1103     & 0.98 \\
$\beta_1$ & 0.1865    & 1.4455     & 0.88 & 0.2760    & 0.1422     & 0.97 & 0.5686    & 0.0646     & 0.99 \\
$\beta_2$ & 0.3677    & 1.8433     & 0.82 & 0.3053    & 0.1593     & 0.98 & 0.7505    & 0.1611     & 0.86 \\
$\beta_3$ & 0.1253    & 1.1679     & 0.87 & 0.1421    & 0.1674     & 0.97 & 0.3440    & 0.0334     & 1.00\\
\hline
\end{tabular}
\caption{$\text{MSE}^*$, MSE and coverage of the posterior 95\% credible interval for the Vague prior, the Logistic prior and the Weighted beta prior.  The summaries refer to 100 samples of size $n=30$.}
\label{T:Freq_3cov_n30}
\end{table}

The MLE averages and the posterior mean averages, are reported in Table \ref{T:Freq_3cov_n30_summaries}.

\begin{table}[h!]
\centering
\begin{tabular}{cccccc}
\hline
          &      &         & \multicolumn{3}{c}{Posterior Means} \\
Parameter & TRUE & MLE     & Vague        & Logistic   & Weighted      \\
\hline
$\beta_0$ & 1.5  & 1.6265  & 1.7309     & 1.5111     & 1.5077    \\
$\beta_1$ & 0.3  & 0.3392  & 0.3748     & 0.2971     & 0.2023    \\
$\beta_2$ & -0.6 & -0.6352 & -0.6782    & -0.5634    & -0.4065   \\
$\beta_3$ & 0.02 & 0.0303  & 0.0533     & 0.0337     & 0.0191   \\
\hline
\end{tabular}
\caption{Average maximum likelihood estimates, and average posterior means for the $n=30$ simulation study.}
\label{T:Freq_3cov_n30_summaries}
\end{table}

%---- n=100 -------
Finally, in Table \ref{T:Freq_3cov_n100} we have reported the results for the larger sample size ($n=100$), where the differences between the priors are reduced, although the better performance of the Logistic and Weighted priors remains.

\begin{table}[h!]
\centering
\begin{tabular}{cccccccccc}
\hline
          & \multicolumn{3}{c}{Vague}       & \multicolumn{3}{c}{Logistic}  & \multicolumn{3}{c}{Weighted}      \\
Parameter & $\text{MSE}^*$ & MSE & Cov  & $\text{MSE}^*$ & MSE & Cov  & $\text{MSE}^*$ & MSE & Cov  \\
\hline
$\beta_0$ & 0.0123    & 0.1647     & 0.92 & 0.0182    & 0.0562     & 0.97 & 0.0259    & 0.0448     & 0.97 \\
$\beta_1$ & 0.0020    & 0.1160     & 0.94 & 0.0054    & 0.0658     & 0.96 & 0.0393    & 0.0396     & 0.98 \\
$\beta_2$ & 0.0025    & 0.0934     & 0.95 & 0.0078    & 0.0547     & 0.99 & 0.0672    & 0.0628     & 0.91 \\
$\beta_3$ & 0.0014    & 0.0952     & 0.97 & 0.0021    & 0.0586     & 0.99 & 0.0144    & 0.0281     & 1.00\\
\hline
\end{tabular}
\caption{$\text{MSE}^*$, MSE and coverage of the posterior 95\% credible interval for the Vague prior, the Logistic prior and the Weighted prior.  The summaries refer to 100 samples of size $n=100$.}
\label{T:Freq_3cov_n100}
\end{table}

The average summaries are reported in Table \ref{T:Freq_3cov_n100_summaries}.

\begin{table}[h!]
\centering
\begin{tabular}{cccccc}
\hline
          &      &         & \multicolumn{3}{c}{Posterior Means} \\
Parameter & TRUE & MLE     & Vague        & Logistic   & Wted      \\
\hline
$\beta_0$ & 1.5  & 1.9354  & 2.5906     & 1.3435     & 1.4266    \\
$\beta_1$ & 0.3  & 0.5153  & 0.7479     & 0.2586     & 0.1092    \\
$\beta_2$ & -0.6 & -0.8578 & -1.1913    & -0.5110    & -0.2388   \\
$\beta_3$ & 0.02 & 0.0833  & 0.1658     & 0.0438     & 0.0154   \\
\hline
\end{tabular}
\caption{Average maximum likelihood estimates, and average posterior means for the $n=100$ simulation study.}
\label{T:Freq_3cov_n100_summaries}
\end{table}

%% file: 6_Application.tex
\section{\label{sec:application}Application to an occupancy model}

 % Rcode in file Induced_Priors_Occupancy_1.Rmd
 % D:\10_Professional_Development\2024_06_Spatial_Occupancy_Modelling_Workshop\Switzerland24-Spatial-Workshop-main\code
 
% Following \citet{2018_Northrup_Gerber} we examined the
% influence of priors on models for occupancy probabilities
% using data on three species of birds (see \citet{1995_McGarigal_McComb}). We use the \texttt{R} package
% \texttt{spOccupancy} \citep{2022_Doser_etal}, in particular
% the function \texttt{PGOcc} is used to
% fit the occupancy models.

We examined the influence of priors on models for presence/absence or occupancy data.  In occupancy modelling Bernoulli random variables appear twice \citep{2017_MacKenzie_etal}, once to indicate the true, but \emph{a priori} unknown, presence (occupancy) of a species at a site or its absence, and again to indicate the detection of the species or failure to detect.  Letting $Z$ be an indicator for true presence and $Y$ be an indicator for detection, the basic occupancy model structure is the following.
 \begin{align*}
    Z & \sim \mbox{Bernoulli}(\psi); \\
    Y|Z=1 & \sim \mbox{Bernoulli}(p),
 \end{align*}
where the probability of detection is here defined conditional on presence. Equivalently, the distribution for $Y$ can be specified simply conditional on $Z$ (rather than $Z=1$), using $Y|Z$ $\sim$ Bernoulli($pZ$). Note also that the unconditional distribution for $Y$ is Bernoulli($p\psi)$. Both occupancy probabilities, $\psi$, and detection probabilities, $p$, are commonly modelled as logistic functions of covariates.  

Here we compare the effects of different priors for the coefficients of the logistic models on a real data set. 
The data are presence/absence counts of European goldfinch made at 266 locations in Switzerland at three different times during the breeding season (Jeffrey Doser, personal communication; see also Chapter 11 of \citet{2015_Kery_Royle} and the \texttt{R} package \texttt{AHMbook} and the dataset MHB2014 \citep{2024_Kery_etal}).  Three covariates were used for modelling occupancy, elevation, elevation$^2$, and the percent forest cover, and three covariates for modelling detection, day-of-year, day-of-year$^2$, and duration of the observation periods (at a given location and visit).   

We use the the function \texttt{PGOcc} in the \texttt{R} package \texttt{spOccupancy} \citep{2022_Doser_etal} to fit the occupancy models.  The default prior in the \texttt{PGOcc} function for the intercept and slope coefficients is Normal(0,1.65$^2$) (labeled spOcc prior), which is a relatively narrow Normal prior compared to vague priors, e.g., Normal(0,1000$^2$). The reason for this choice is partially based on results from \citet{2018_Northrup_Gerber} (Jeffrey Doser, personal communication). We used the same logistic prior for the intercept and the slope coefficients as in Section \ref{sec:mean.var.determination}. This was done largely for simplicity as \texttt{PGOcc} is set up for using the same priors for the intercepts and slopes (allowing for differences between occupancy probability and detection probability).  Given $p$=3 covariates, the logistic model priors were then Normal(0, 0.91$^2$).  A third set of priors labelled wide priors were also used, Normal(0, 40$^2$), instead of the Vague priors used previously, Normal(0, 1000$^2$), which caused \texttt{PGOcc} to fail. 

\begin{figure}[h]
    \centering
    \includegraphics[width=0.7\linewidth]{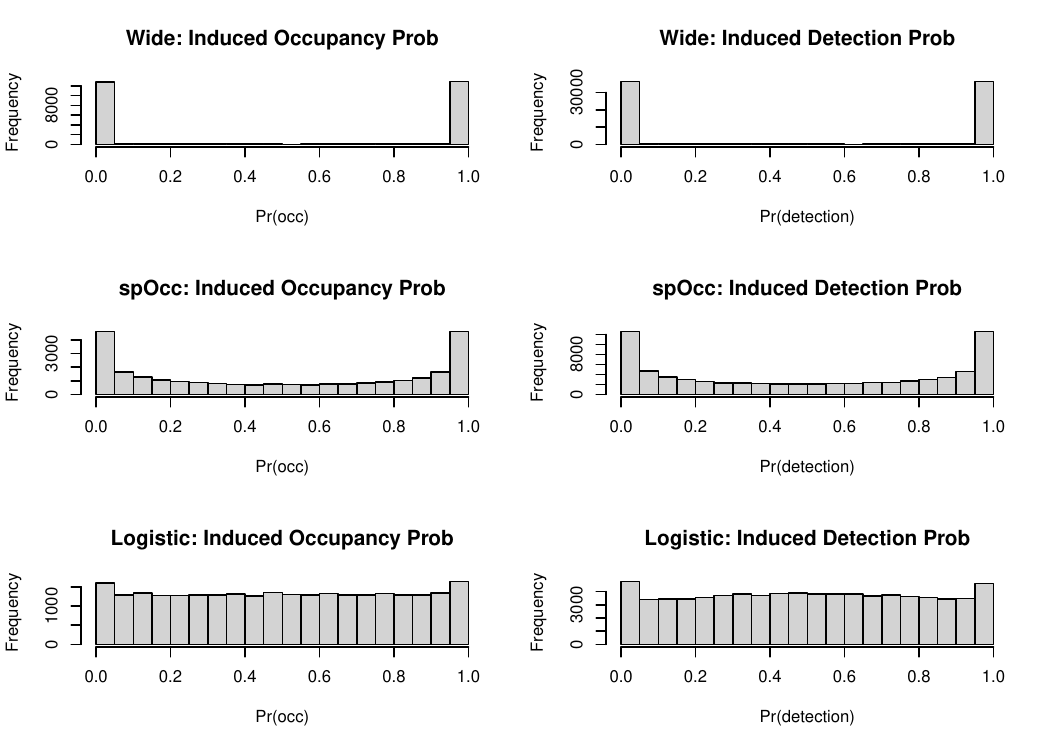}
    \caption{Induced prior distributions for occupancy and detection probabilities of the goldfinch data set for three different priors for the logistic regression intercept and partial slopes.}
    \label{F:induced.priors.goldfinch}
\end{figure}

\begin{figure}[h]
    \centering
    \includegraphics[width=0.7\linewidth]{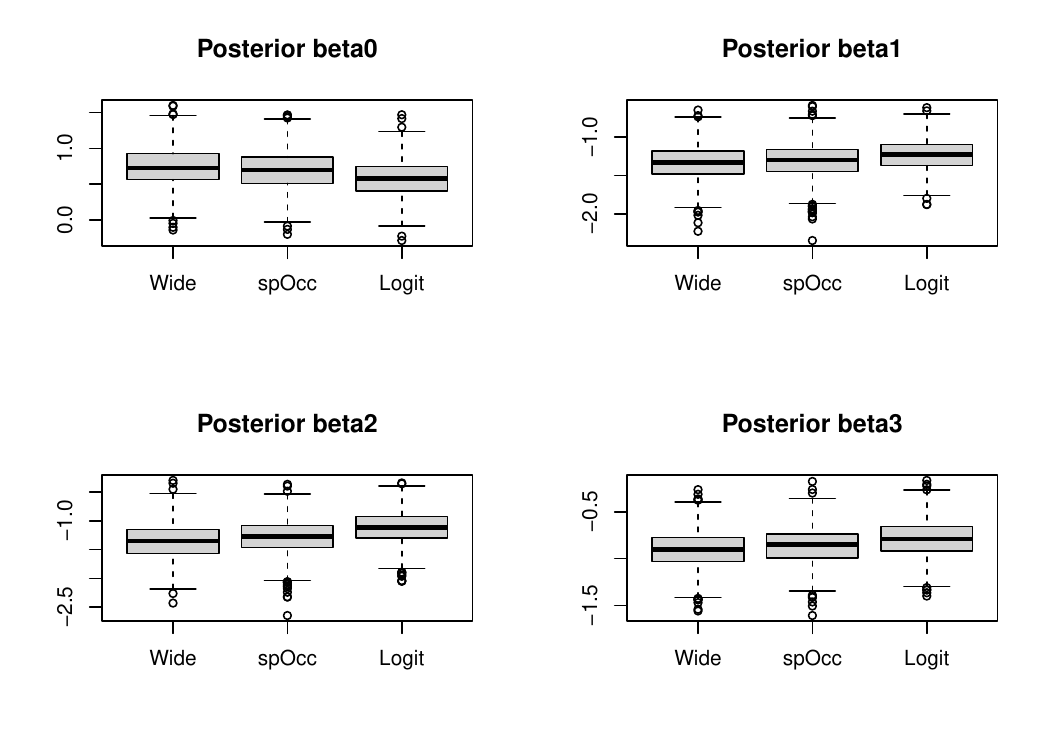}
    \caption{Posterior distributions for Occupancy model intercept and slope coefficients for three different priors.}
    \label{F:posterior.All.occupancy.params}
\end{figure}

\begin{figure}[h]
    \centering
    \includegraphics[width=0.7\linewidth]{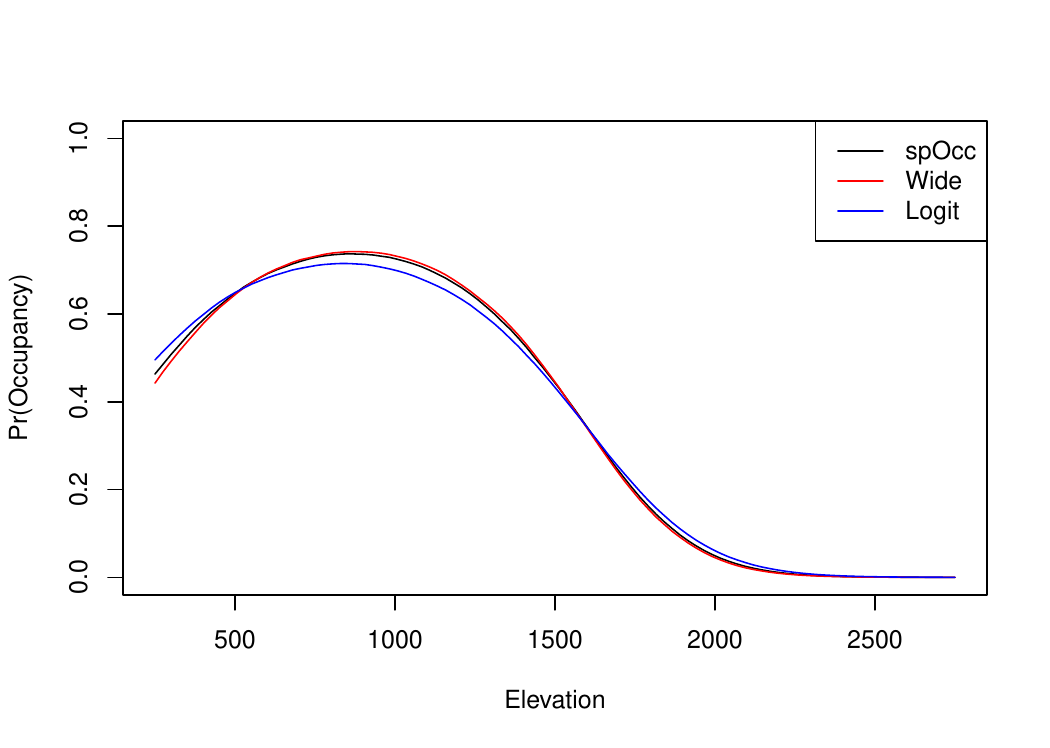}
    \caption{Posterior means for the probability of occupancy as a function of elevation (with percent forest cover set at average value) for the three different priors.}
    \label{F:posterior.All.occupancy.prob}
\end{figure}

The usual bathtub shape of the induced priors for the occupancy and the detection probabilities was observed for both the wide and  spOcc priors, while the induced priors from the logistic model priors were relatively flat (Figure \ref{F:induced.priors.goldfinch}). The resulting posterior distributions for the intercepts and slopes differed slightly between the three different priors, for example, the posterior distributions for the occupancy coefficients (Figure \ref{F:posterior.All.occupancy.params}) with the logistic prior yielding posterior distributions for the slope coefficients shifted slightly more positive than the other two priors. The resulting posterior mean probabilities for occupancy as a function of elevation, shown in Figure \ref{F:posterior.All.occupancy.prob}, were relatively similar across the range of elevations.  The posterior credible intervals (not shown) were relatively wide, particular at the extremes of elevation which had less data.

\begin{figure}[h]
    \centering
    \includegraphics[width=0.7\linewidth]{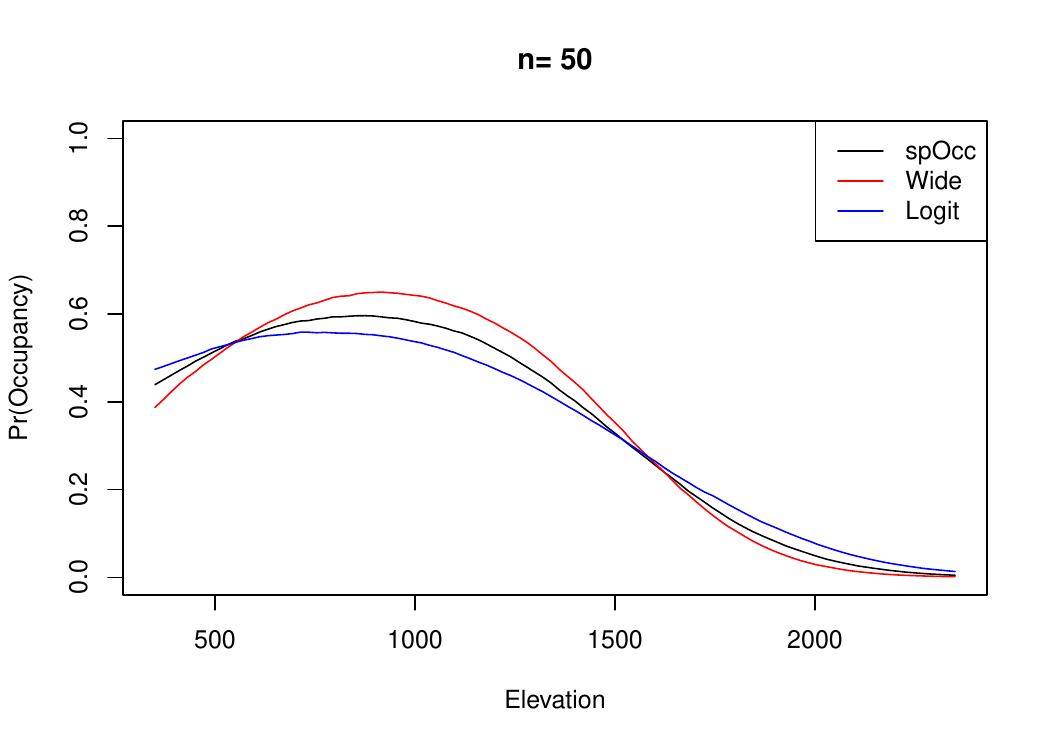}
    \caption{Resulting posterior means for the probability of occupancy as a function of elevation for a subset of the $n$=50 locations in the goldfinch occupancy survey.}
    \label{F:posterior.reduced.50.occupancy.prob}
\end{figure}

The relative similarity in the posterior distributions can be attributed to the influence of the data overshadowing the influence of the prior.  This can be seen in Figure \ref{F:posterior.reduced.50.occupancy.prob} which shows the resulting posterior mean occupancy probability as a function of elevation when the model was fit using a simple random sample of $n$=50 drawn from the 266 survey locations. The discrepancy between the three different priors is more pronounced with $n$=50 than when all $n$=266 sites were used, with the spOcc  and logistic priors yielding somewhat similar results but the wide prior less so. 

\clearpage 

%% file: 7_Discussion_Conclusions.tex
\section{\label{sec:Discussion}Discussion}

We have proposed priors for the coefficients in a multiple covariate logistic regression for a Bernoulli probability $\theta$ that induced a desired prior for $\theta$. To the best of our knowledge previous work has focused largely on single covariate models or has applied the same prior for each regression coefficients irregardless of the number of covariates.  The proposed priors include ones such that the induced prior for $\theta$ is an  Beta($\alpha, \beta$) with arbitrary shape parameters as well as priors that allow for distinct handling of the intercept in the regression model.  This work has raised additional questions that suggest other areas of research that are included below.

\subsection{Weighting in the Weighted prior}
For the Weighted prior, the proposed solution for determining the variance for the intercept was to set it equal to some fraction of $\sigma^2_\eta$, $\sigma^2_0$ = $k \sigma^2_\eta$, $0 \le k \le 1$. Can some guidance be provided regarding the choice of $k$?  

If $k$ is near 0, then the prior for the intercept will be concentrated around $\mu_\eta$, and the priors for the slope coefficients will be near their maximum dispersion, $\sigma^2_\eta/p$. At the extreme of $k$=0, the intercept $\beta_0$ is a fixed value.  On the other hand if $k$ is near 1, then the variance of the prior for the intercept is near its maximum $\sigma^2_\eta$ and the priors for the slope coefficients are more concentrated around zero. At the extreme of $k$=1 all slope coefficients disappear. 

Therefore $k$ is balancing the effect of the intercept and the covariates. This begs the question of how to choose $k$. One thought is to treat $k$ as a parameter to estimate and then specify a hierarchical prior distribution on it.  Another is to view $k$ as a tuning parameter in terms of a goodness-of-fit measure. Further investigation of this issue is a topic of further research. 

\subsection{Concentration of priors around 0}
As the number of covariates increases, the priors for the slope coefficients are increasingly concentrated  around 0, i.e., there is shrinkage toward 0.   Two statistical procedures that have some similarity with this result is Lasso \citep{1996_Tibshirani} and penalized complexity priors \citep{2017_Simpson_etal}. Lasso can cause shrinkage of coefficients to the point that a covariate is completely removed. Penalized complexity priors aim, given a set of nested models, to give highest prior probability to a ``base'' model, e.g., an intercept only model.  Whether or not this shrinkage is desireable and whether or not more direct connections can be made with Lasso or penalized complexity priors are additional areas for research.

\subsection{Generating functions}
The linear combination of random variables, namely the intercept and partial slope coefficients in the linear model, is a convolution.  Dropping the intercept and ignoring the effects of the covariate values in the linear combination,  assuming all covariates equal 1, if identical distributions are assumed for the slope coefficients, which is a way to address the many-to-one problem mentioned previously,  the moment generating function (MGF) for the linear combination is the product of $p$ identical MGFs:
\begin{align*}
E[\exp(\eta)] &= E[\exp(t \times [\beta_1  + \ldots + \beta_p ])] =
 E[\exp(t \beta)]^p,
\end{align*}
where $\beta$, $\beta_1$, $\ldots$, and $\beta_p$ are independent and identically distributed. This results in the MGF for $\beta$,
\begin{align*}
 E[\exp(t \beta)] &= 
 \left ( E[\exp(\eta)] \right )^{\frac{1}{p}}.
\end{align*} 
If the MGF for $\eta$ is known, the MGF for $\beta$ the $p$th root of that MGF.  Inversion of an arbitrary MGF to yield the underlying probability distribution is usually not a simple task let alone inversion of the $p$th root of a MGF \citep{2017_Walker}. While interesting, this approach seems impractical.

\subsection{Multi-Bernoulli distributions}
Another topic for research is to extend these results to multi-Bernoulli distributions.  For example, an animal is
at location $A$ at time $t$ and at time $t+1$ it could  
remain  at $A$ or move  to $B$ or move  to $C$. Model location
at next time step by a multi-Bernoulli:
\begin{align*}
y_{t+1}|y_t & \sim \mbox{Mult-Bernoulli}(p_1, p_2, p_3).
\end{align*}
The probabilities are modelled by a multi-logistic transformation:
\begin{align*}
    p_1 & \propto  \exp(\beta_{1,0} + \beta_{1,1}x_1) \\
    p_2 & \propto \exp(\beta_{2,0} + \beta_{2,1} x_2) \\
    p_3 & = 1-p_1-p_2.
\end{align*}
If discrete
uniform priors for $p_1$, $p_2$, and $p_3$ are desired, what do the priors for the above $\beta$'s need to be? 
 
\subsection{General problem of lack of parameterization invariance}
 Here we have only addressed potentially problematic induced priors for Bernoulli parameters.  Both \citet{2012_Seaman_etal} and \citet{2020_Lele} discussed this problem for non-Bernoulli settings.  In particular \citet{2020_Lele} compared two mathematically equivalent parameterizations for a population dynamics model in a state-space model framework, namely the Ricker model.  The two formulations are
 \begin{align}
 \label{eq:Ricker.A}
\mbox{Model A  :} N_{t+1} &= N_t \times \exp(a-b \times N_t) \\
\label{eq:Ricker.B}
\mbox{Model B  :} N_{t+1} &= N_t \times \exp(a \times (1-N_t/K)),
 \end{align}
 where $a$ is a measure of intrinsic population growth rate, $b$ is a measure of density dependence, and $K$ is carrying capacity. $K$ can be calculated for the first model as $a/b$.
 He showed how apparently non-informative priors for the parameters yielded quite different posterior means for the population growth parameter and population carrying capacity. 

\begin{figure}[h]
     \centering
     \includegraphics[width=0.5\linewidth]{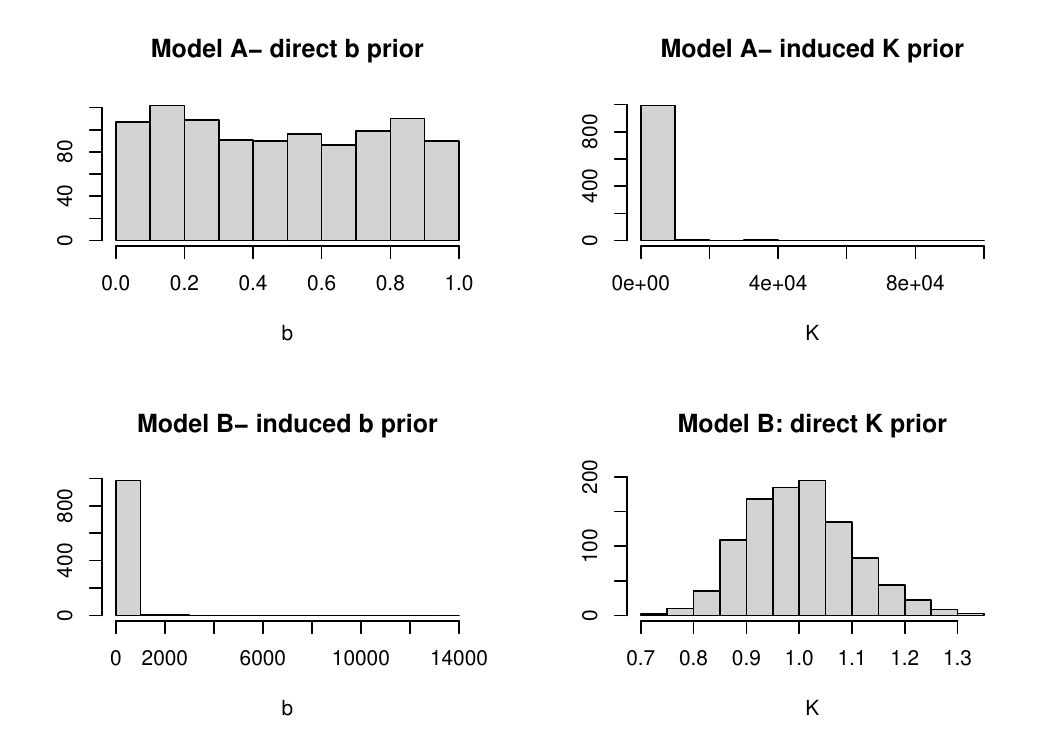}
     \caption{Priors, explicit and induced, for two parameterizations of the Ricker model used by \citet{2020_Lele}. Model A formulation is Equation (\ref{eq:Ricker.A}) and Model B formulation is Equation (\ref{eq:Ricker.B}).}
     \label{F:Ricker.induced.priors}
 \end{figure}

 One likely explanation for the discrepancy in the posterior means is that the induced priors for the functions of parameters with specified priors are quite different than the explicitly specified priors for those parameters in the other parameterization.  Figure \ref{F:Ricker.induced.priors} shows how the induced distribution for carrying capacity $K$ in Model A is highly skewed compared to specified distribution for $K$ in Model B, with a similar result for the induced prior for the growth parameter $a$ in Model B compare to the specified prior in Model A.
 
 In principle the same concepts used to yield a desired induced prior distribution for the Bernoulli parameter could be applied by specifying desired induced priors for the growth rate  and the carrying capacity parameters---using the change-of-variables approach to work backwards to derive the prior probabilities on the lower level parameters that would induce the desired priors.  However, the resulting derived prior distributions will typically be quite unique and Monte Carlo procedures will likely be required to generate samples from the probability function.  
 
%   Count data. $Y|X$ is Poisson($\lambda(X)$) and $\ln(\lambda(X))$ is modelled as $\beta_0 + \beta_1 X$.
% Suppose desired induced prior distribution of $\lambda(\overline{X})$ is D(?,?).

%% file: Appendix.tex
\section*{\label{app:sim.studies}Appendix - Simulation studies details}
Additional details and output from the simulations discussed in Section \ref{sec:simulations} are given below. 

%-------------------------------------------------------
%-------------------------------------------------------
\subsection*{Scenario 1: One covariate and $\theta\sim U(0,1)$}

{\bf Single Sample Analysis}

\begin{figure}[h]
     \centering
     \includegraphics[width=0.5\linewidth]{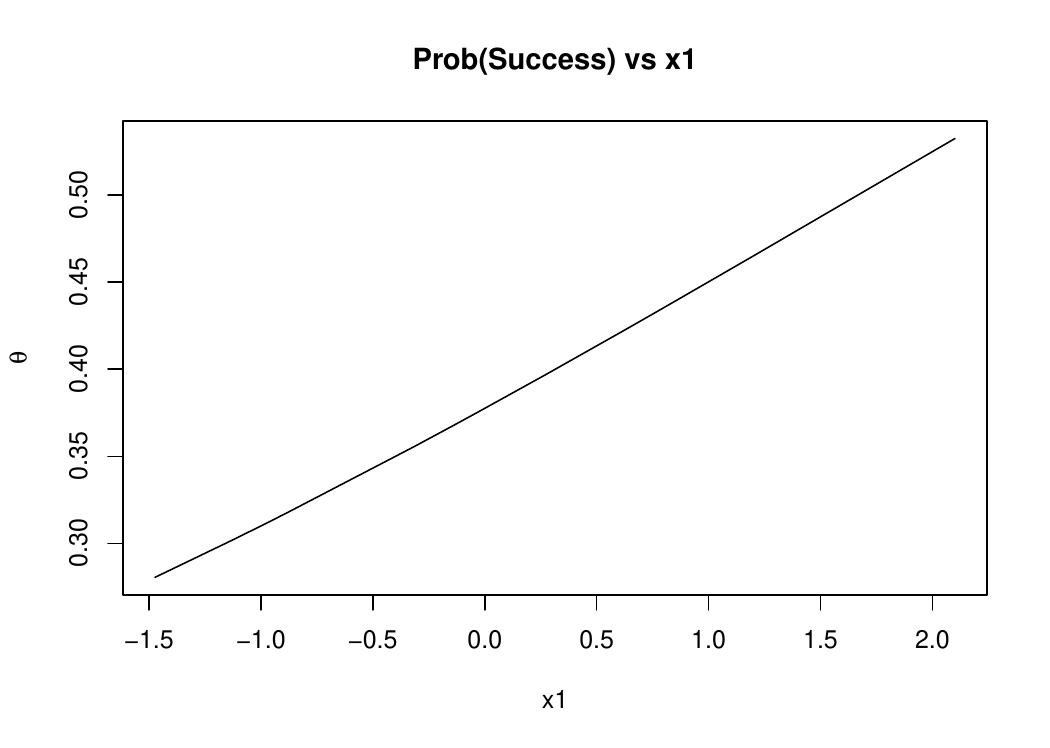}
     \caption{Scenario 1 logit model: $\theta$ vs $x$.}
     \label{F:univariate.theta}
 \end{figure}

We have drawn a sample of size $n=15$ from the model given in Equation \eqref{eq:model_scenario_1} in Section \ref{sec:scenario1}. The simulated covariate values and related responses are shown in Table \ref{T:univariate.data}, while the relationship between $\theta$ and $x$ is plotted in Figure \ref{F:univariate.theta}.

\begin{table}[ht]
 \centering
 {\scriptsize 
 \begin{tabular}{rrr}
   \hline
  & y & x1.std \\ 
   \hline
 1 & 1.00 & -1.47 \\ 
   2 & 0.00 & -1.13 \\ 
   3 & 1.00 & -1.01 \\ 
   4 & 0.00 & -1.01 \\ 
   5 & 1.00 & -0.90 \\ 
   6 & 1.00 & -0.31 \\ 
   7 & 1.00 & -0.15 \\ 
   8 & 1.00 & -0.03 \\ 
   9 & 0.00 & 0.26 \\ 
   10 & 1.00 & 0.28 \\ 
   11 & 1.00 & 0.61 \\ 
   12 & 1.00 & 0.73 \\ 
   13 & 1.00 & 0.84 \\ 
   14 & 1.00 & 1.19 \\ 
   15 & 1.00 & 2.10 \\ 
    \hline
 \end{tabular}
 }
 \caption{Simulated responses and covariates
 for single covariate model.}
 \label{T:univariate.data}
 \end{table}

\begin{figure}[h]
    \centering
    \includegraphics[width=0.6\linewidth]{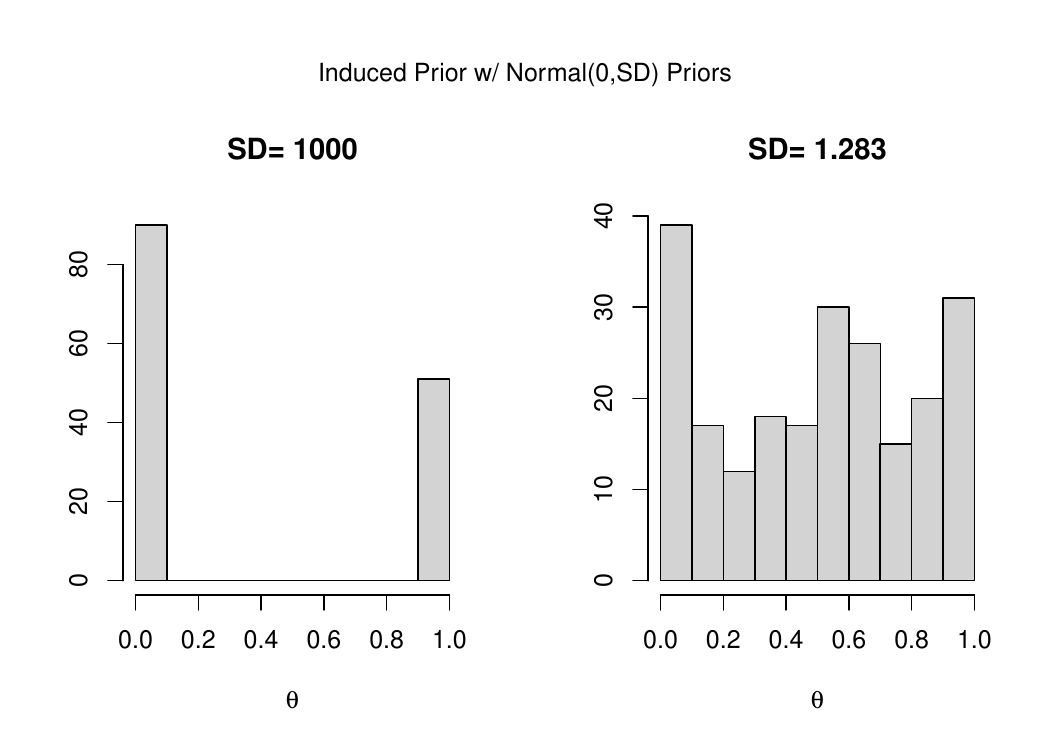}
    \caption{Induced priors for $\theta$ based
    on a single covariate with Normal(0, 1000$^2$)
    and Normal(0, $\frac{\pi^2}{3 \times 2})$ priors based on $n$=15 observations.}
    \label{F:induced.theta.priors.univariate}
\end{figure}

The induced priors on $\theta$ obtained by implementing the priors on the $\beta$s (see Equation \eqref{eq:prior_scenario_1}), can be found in Figure \ref{F:induced.theta.priors.univariate}. We note that there is a difference in shape between the two priors: the one based on the Vague prior assigns almost all the mass towards 0 and 1, whilst the induced prior resulting from the method we proposed can be approximated by a uniform density (which was the aim).

\begin{figure}[h]
    \centering
    \includegraphics[width=0.5\linewidth]{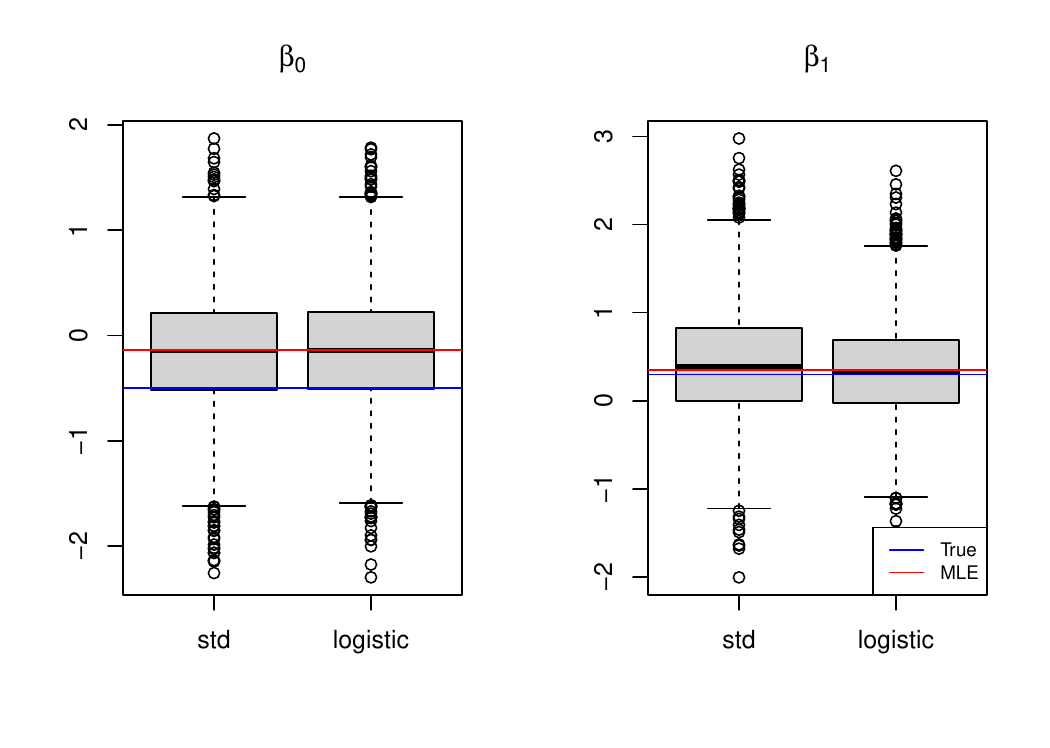}
    \caption{Posterior distributions for $\beta_0$ and 
    $\beta_1$ with Vague and Logistic priors.}
    \label{F:post.univariate}
\end{figure}

\begin{figure}[H]
    \centering
    % Top subfigure (4 plots in 2x2 grid)
    \begin{subfigure}[b]{\textwidth}
        \centering
        \begin{subfigure}[b]{0.40\textwidth}
            \centering
            \includegraphics[width=\textwidth, scale=1.0]{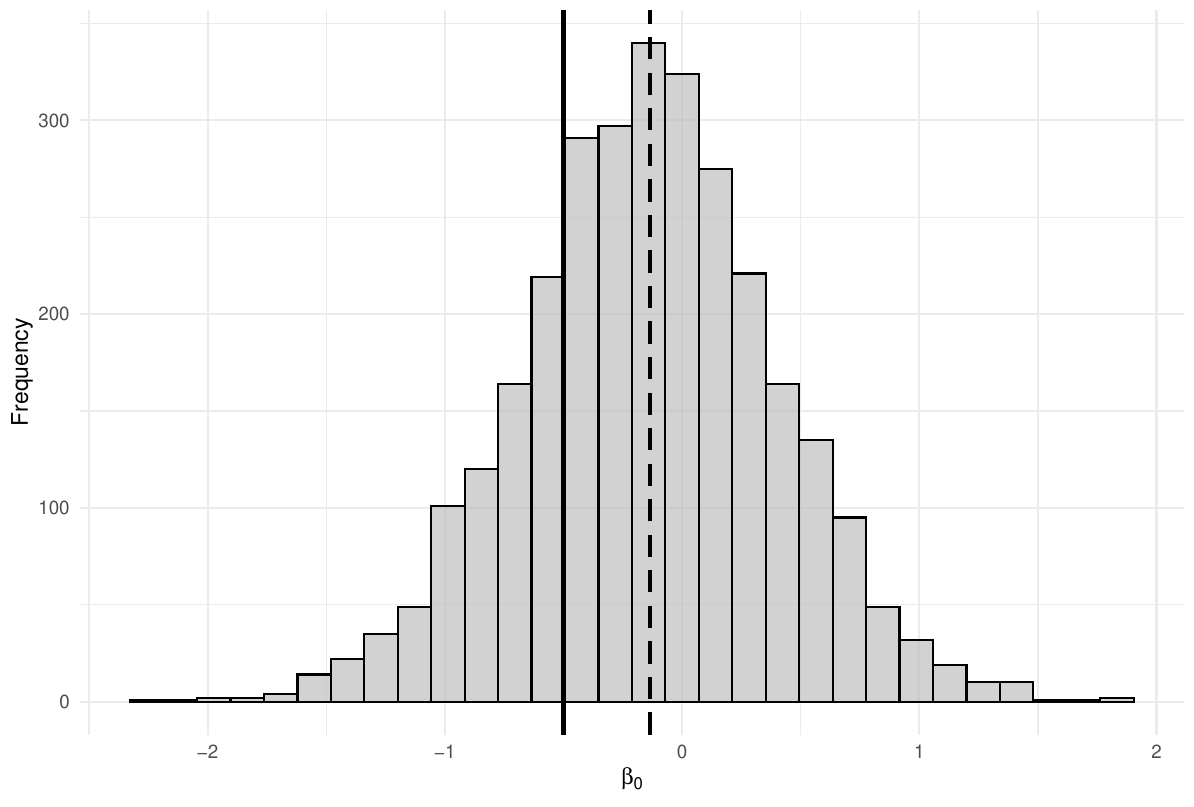}
            %\caption{Plot 1}
        \end{subfigure}
        \hfill
        \begin{subfigure}[b]{0.40\textwidth}
            \centering
            \includegraphics[width=\textwidth, scale=1.0]{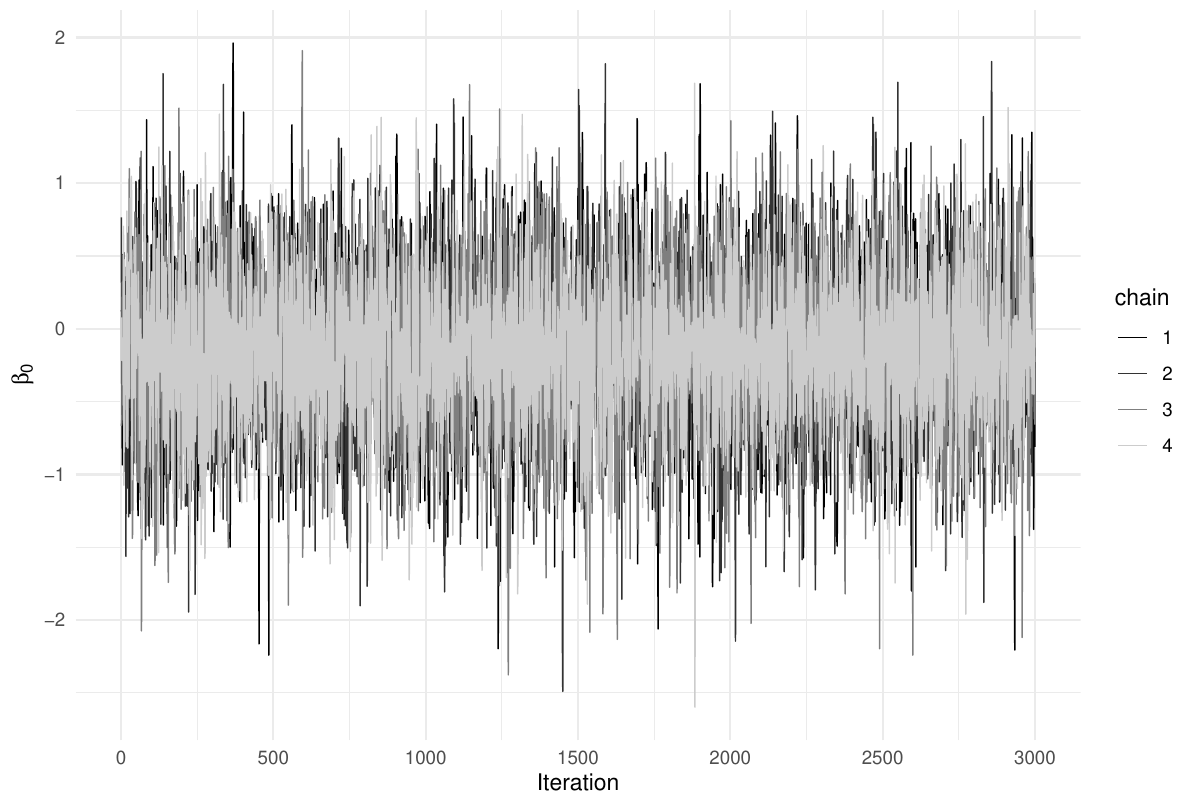}
            %\caption{Plot 2}
        \end{subfigure}
        
        %\vspace{0.5cm}  % Space between rows
        \vspace{0.1cm}  % Space between rows
        
        \begin{subfigure}[b]{0.40\textwidth}
            \centering
            \includegraphics[width=\textwidth, scale=1.0]{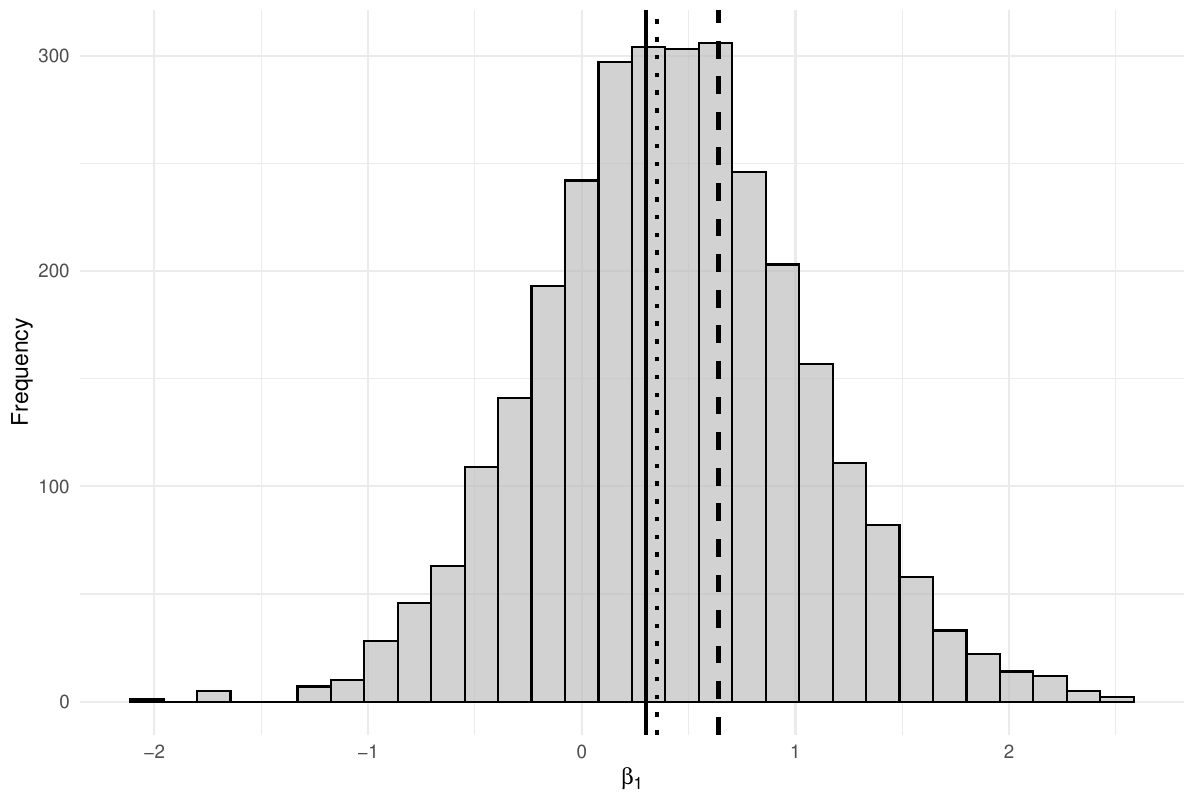}
            %\caption{Plot 3}
        \end{subfigure}
        \hfill
        \begin{subfigure}[b]{0.40\textwidth}
            \centering
            \includegraphics[width=\textwidth, scale=1.0]{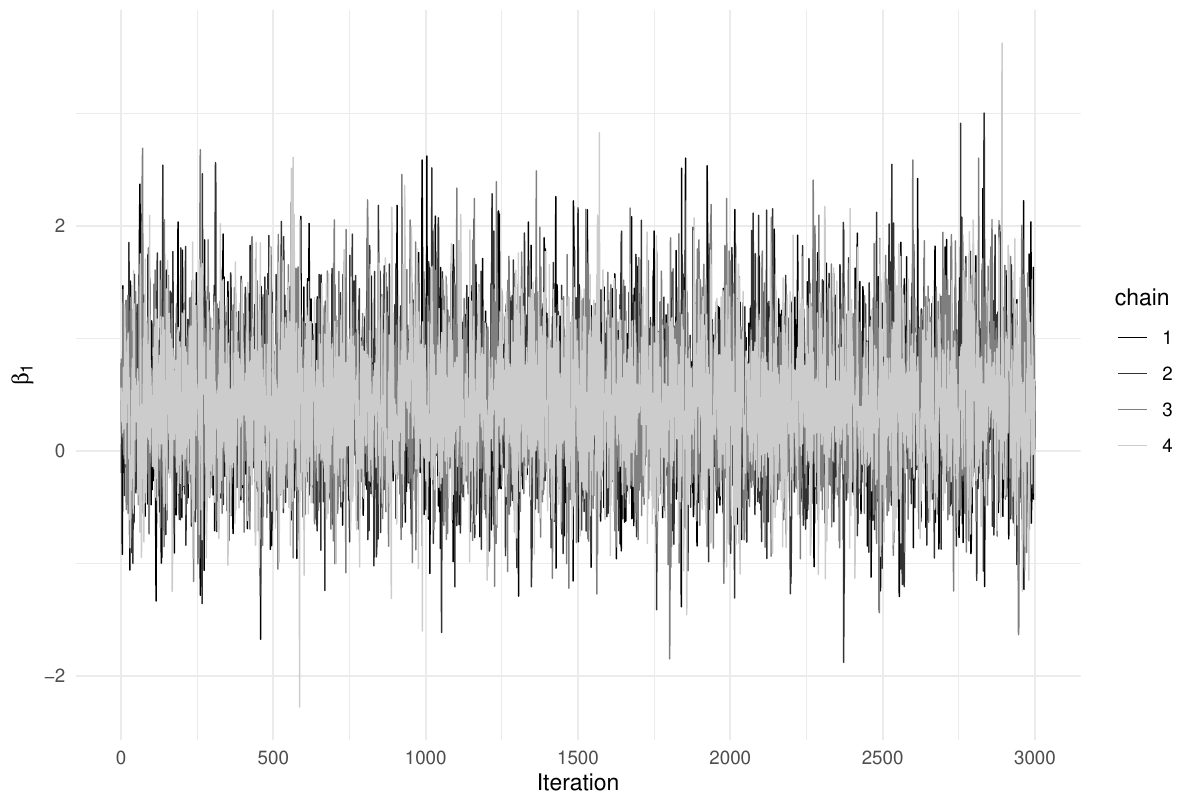}
            %\caption{Plot 4}
        \end{subfigure}
        \caption{ }
    \end{subfigure}
    
    %\vspace{1cm}  % Space between top and bottom subfigures
    \vspace{0.2cm}  % Space between top and bottom subfigures
    
    % Bottom subfigure (4 plots in 2x2 grid)
    \begin{subfigure}[b]{\textwidth}
        \centering
        \begin{subfigure}[b]{0.40\textwidth}
            \centering
            \includegraphics[width=\textwidth]{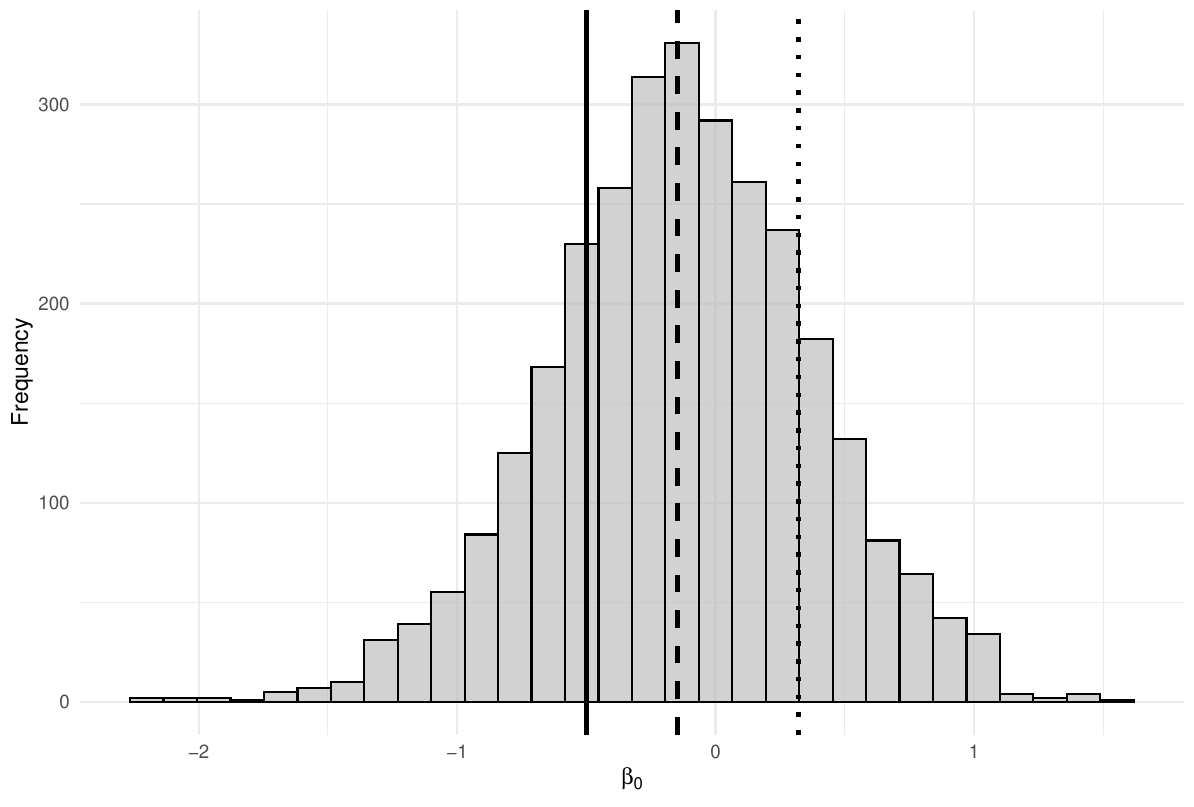}
            %\caption{Plot 5}
        \end{subfigure}
        \hfill
        \begin{subfigure}[b]{0.40\textwidth}
            \centering
            \includegraphics[width=\textwidth]{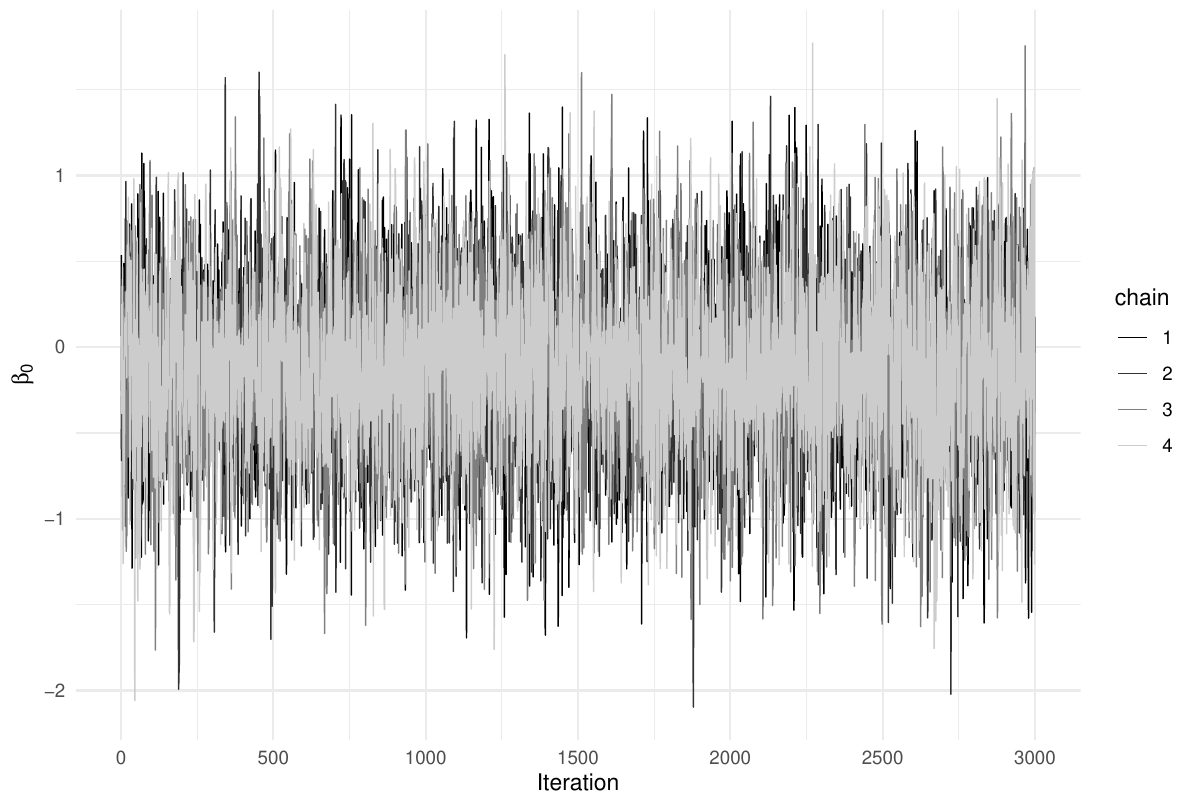}
            %\caption{Plot 6}
        \end{subfigure}
        
       %\vspace{0.5cm}  % Space between rows
        \vspace{0.1cm}  % Space between rows
         
        \begin{subfigure}[b]{0.40\textwidth}
            \centering
            \includegraphics[width=\textwidth]{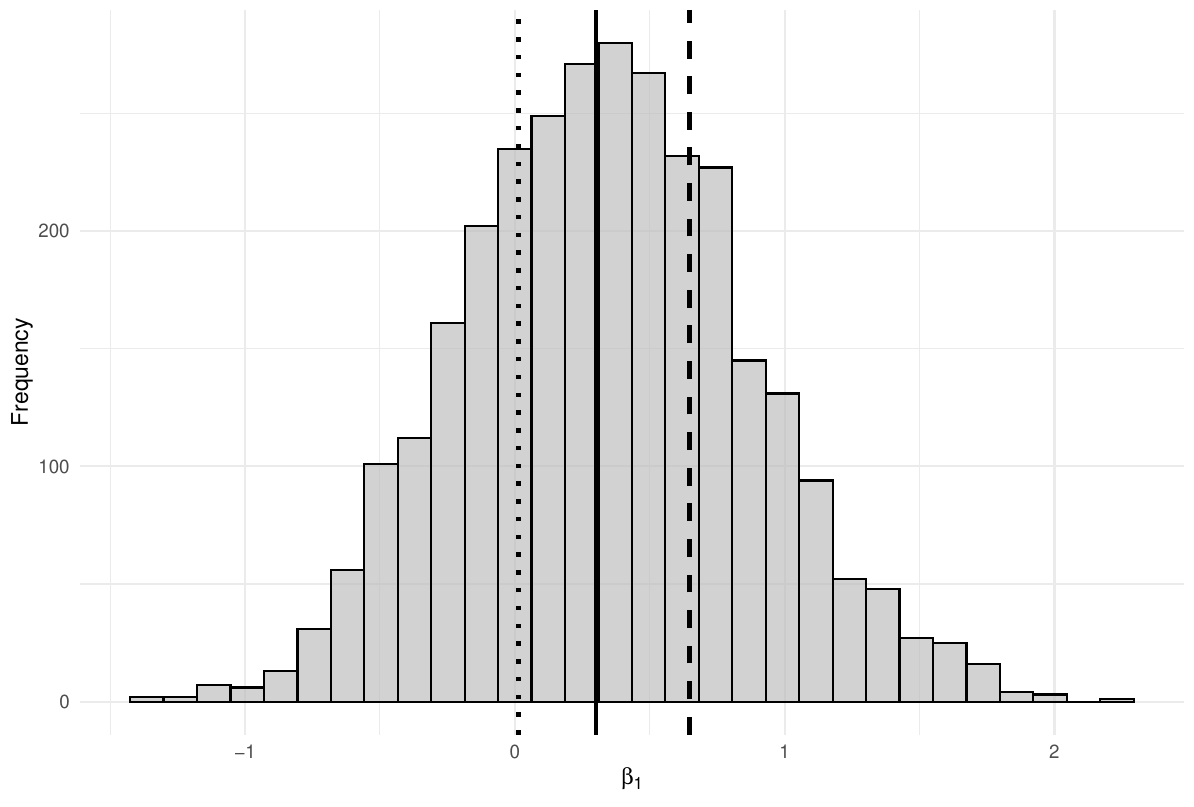}
            %\caption{Plot 7}
        \end{subfigure}
        \hfill
        \begin{subfigure}[b]{0.40\textwidth}
            \centering
            \includegraphics[width=\textwidth]{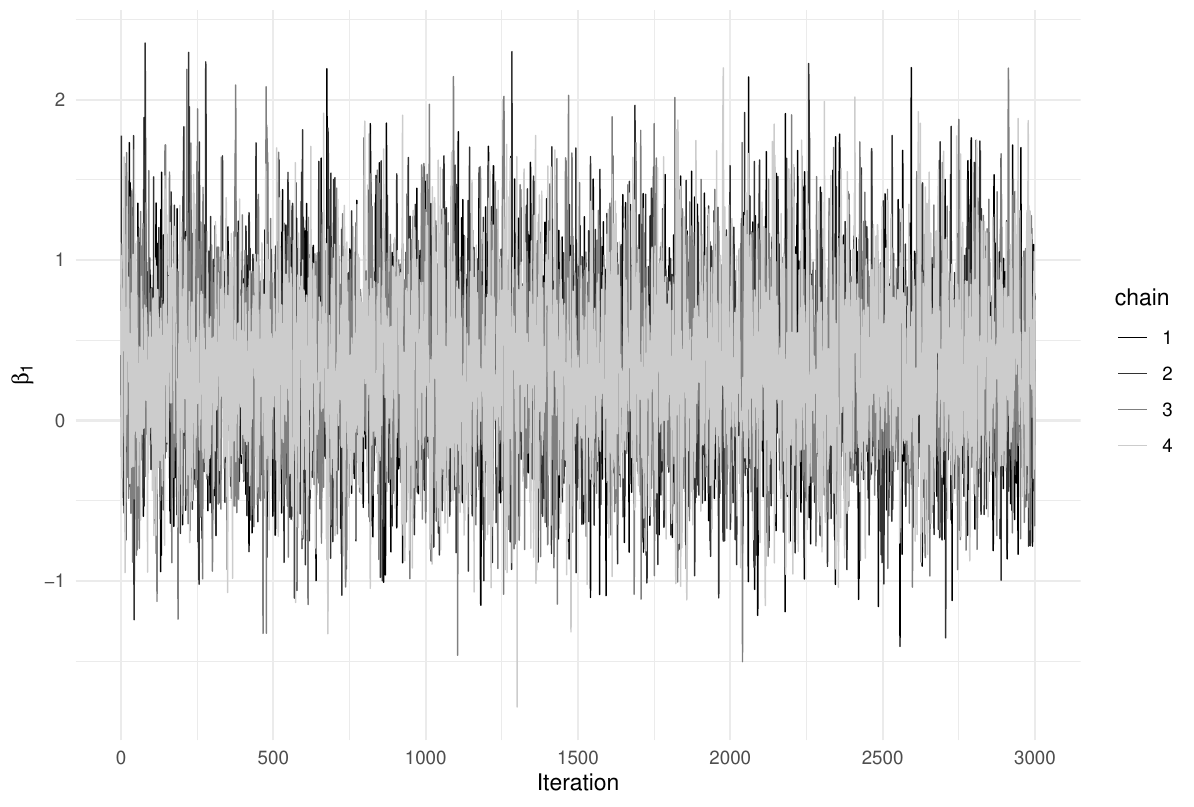}
            %\caption{Plot 8}
        \end{subfigure}
        \caption{ }
    \end{subfigure}
    \caption{Scenario 1: Posterior histograms and chains of the intercept and the  slope coefficient for the (a) vague prior and the (b) proposed prior. Each histogram contains three vertical lines showing the true value (solid), the MLE (dotted) and the posterior mean (dashed). We have sampled $n=15$ data points from a model with parameters $\beta_0=-0.5$ and $\beta_1=0.3$.}
    \label{F:SingleSample_scenario 1A}
\end{figure}

Side-by-side boxplots of the posterior distributions for $\beta_0$ and $\beta_1$ resulting from the Vague and Logistic priors are shown in Figure \ref{F:post.univariate}, while the histograms of the posterior distributions for the intercept and the slope coefficient, under both priors, along with the MCMC trace plots, are shown Figure \ref{F:SingleSample_scenario 1A}. The MCMC was run for 5000 simulations with a burnin of 2000 iterations using 4 independent chains. No evidence of a lack of convergence was identified. 

 In Table \ref{T:Scenario1.single.dataset.results} we show the posterior summaries, that is the posterior mean and the posterior 95\% credible interval, as well as the true value and the MLE. We see that the true parameters are contained in both posterior 95\% credible intervals, and the intervals are larger under the Logistic prior, meaning that it carries less information than the Vague prior. The result, therefore, supports our statement of being a ``more objective'' prior. Although the posterior distributions are reasonably symmetric, as seen in Figure \ref{F:SingleSample_scenario 1A}, the maximum a posteriori (MAP) are -0.13 and -0.15 for the intercept, under the Vague and Logistic prior, respectively, and 0.64 and 0.65 for the slope coefficient. The last results suggest negatively skewed posteriors.

\begin{table}[h!]
    \centering
    \begin{tabular}{ccccccc}
    \hline
    % First row with merged columns
        & & & \multicolumn{2}{c}{Vague} & \multicolumn{2}{c}{Logistic}\\
        Parameters & Truth & MLE & Mean & C.I. & Mean & C.I. \\
        \hline
        $\beta_0$ & -0.50 & -0.14 & -0.13 & (-1.28, 0.94) & -0.13 & (-0.74, 1.67) \\
        $\beta_1$ & 0.30 & 0.35 & 0.43 & (-1.13, 0.86) & 0.34 & (-0.70, 1.43) \\
        \hline
    \end{tabular}
    \caption{Scenario 1: Posterior mean and posterior 95\% credible intervals with $n=15$. The table includes also the true values and the MLEs.}
    \label{T:Scenario1.single.dataset.results}
\end{table}

\subsection*{Scenarios 2 and 3}
%\subsubsection{\label{sec:scenario2.single}Single dataset}

{\bf Single Sample Analysis}

\begin{figure}[h]
    \centering
    \includegraphics[width=0.5\linewidth]{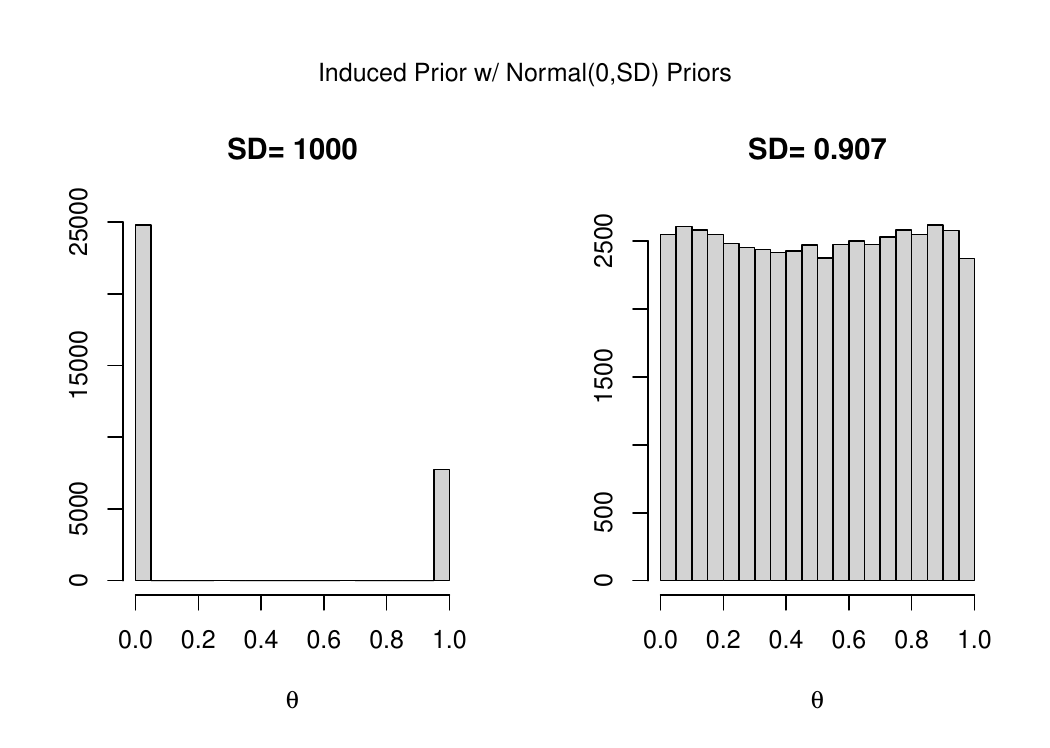}
    \caption{Induced $\theta$ priors with
    3 covariates with coefficient priors of
      Normal(0,1000$^2$)
    and Normal
    $\left   (0, \frac{\pi^2}{3 \times (1+3)}=0.907^2  \right   )$ 
    priors based on $n$=50 observations.}
    \label{F:induced.theta.priors.multivariate}
\end{figure}

\begin{figure}[h]
    \centering
    \includegraphics[width=0.5\linewidth]{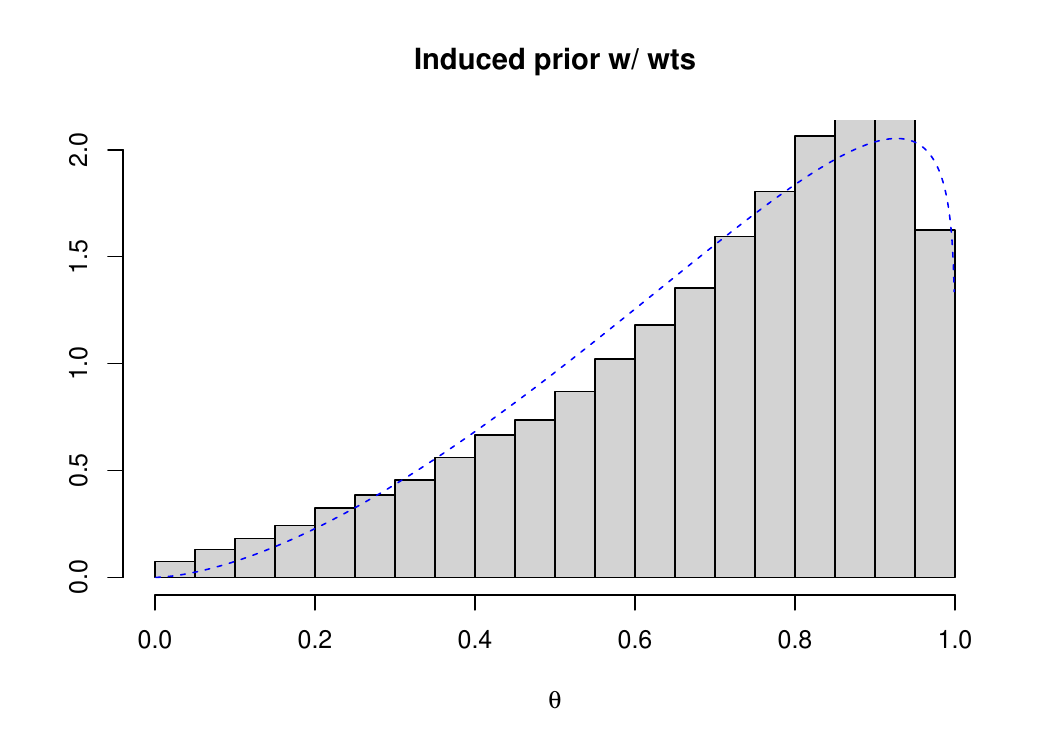}
    \caption{Histogram of induced prior for $\theta$ given
    Weighted priors for intercept and slope
    coefficients along with Beta(2.633, 1.129)
    pdf (in blue dashed lines.)}
    \label{F:induced.wted.prior.theta}
\end{figure}

For this experiment, we simulated a sample of size $n=50$ from the model given in Equation \eqref{eq:multiple.covariates} in Section \ref{sec:scenario2}. The prior distributions on the $\beta$s were the Vague prior, the Logistic prior and the Weighted prior (see Equations \eqref{eq:prior_scenario_2} and \eqref{eq:weightedprior}). The induced priors on $\theta$ corresponding to the Vague and the Logistic prior are shown in Figure \ref{F:induced.theta.priors.multivariate}, while the one for the weighted prior is shown in Figure \ref{F:induced.wted.prior.theta}. We recall that the Weighted prior is designed so that the induce prior on $\theta$ is an arbitrary $\text{Beta}(\alpha,\beta)$ and, in addition, the prior for the intercept $\beta_0$ is distinct from the one on the coefficients. In our case, the prior distribution for $\theta$ was determined by
specifying an expected value of 0.7 and a CV=0.3, which leads to Beta(2.633, 1.129).   The induced expectation for $\eta$ was $\mu_\eta$=1.150 and $\sigma^2_\eta$ = 1.843. 
% E[eta]= 1.150069 Var[eta]= 1.842635
A weight for the variance for the prior for $\beta_0$ of $k$=0.4 was selected. The resulting prior is Equation \eqref{eq:weightedprior} in Section \ref{sec:scenario2}.

The MCMC parameters are as in the previous example, that is we run 4 chains for 5000 simulations and a burnin of 2000 iterations, giving a usable posterior sample of 3000 iterations.
The posterior summaries are reported in Table \ref{T:SingleSample_3cov_small}, and the  histograms of the posterior distributions are in Figure \ref{F:SingleSample_scenario_2+3cov} in Appendix \ref{app:sim.studies}. Given that some of the posterior distributions appear to be asymmetric, we have also reported the MAPs in Table \ref{T:map_n50_3cov_small}, along with the corresponding true values and MLEs.

The prior performances vary across the parameters, although the true values are all within the 95\% posterior credible intervals. There no clear dominance of a prior over the others in terms of ``guessing'' the true value nor in getting closer to the MLE considering as point estimate the posterior mean.
\begin{table}[h!]
\centering
\begin{tabular}{ccccccccc}
\hline
         & &       & \multicolumn{2}{c}{Vague} & \multicolumn{2}{c}{Logistic} & \multicolumn{2}{c}{Wted} \\
Parameter & Truth & MLE & Mean  & C.I.            & Mean      & C.I.             & Mean    & C.I.           \\
\hline
$\beta_0$      & 1.50 & 1.80 & 2.05  &      (1.19, 3.07)  & 1.59    & (0.91, 2.33)  & 1.65  & (1.01, 2.35)   \\
$\beta_1$      & 0.30 & 0.11 & 0.12  &      (-0.83, 1.09) & 0.08    & (-0.67, 0.84) & 0.04  & (-0.49, 0.57)  \\
$\beta_2$      & -0.60 & -0.62 & -0.70 &    (-1.63, 0.15) & -0.49   & (-1.22, 0.21) & -0.29 & (-0.82, 0.24)  \\
$\beta_3$      & 0.02 & 0.27 & 0.37 &       (-0.56, 1.43) & 0.26    & (-0.45, 1.04) & 0.15  & (-0.38, 0.70) \\
\hline
\end{tabular}
\caption{Posterior means and 95\% credible intervals for the Vague, Logistic, and Weighted priors ($n$=50).}
\label{T:SingleSample_3cov_small}
\end{table}

\begin{table}[h]
    \centering
    \begin{tabular}{cccccccc}
    \hline
  Parameters & True & MLE & Vague & Logistic & Wted \\
        \hline
  $\beta_0$ & 1.50    & 1.80  & 2.24  & 1.49  & 1.75 \\
  $\beta_1$ & 0.30    & 0.11  & 0.20  & -0.06 & 0.24 \\
  $\beta_2$ & -0.60   & -0.70 & -1.01 & -0.60 & -0.13 \\
  $\beta_3$ & 0.02    & 0.37  & 0.27  & 0.04  & 0.04 \\
  \hline
  \end{tabular}
 \caption{MAP under the Vague, Logistic, and Weighted priors for the case $n=50$.}
    \label{T:map_n50_3cov_small}
\end{table}
% Again, even considering the MAPs, instead of the posterior means, we are not able to identify a prior distribution that outperforms the others for all the parameters.

\begin{figure}[h]
    \centering
    % First row
    \begin{subfigure}[b]{0.3\textwidth}
        \centering
        \includegraphics[width=\textwidth]{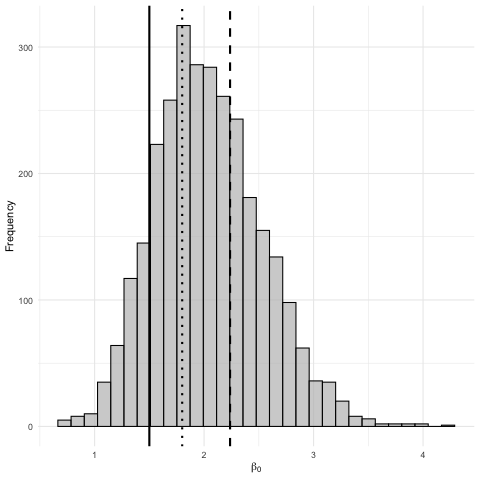}
        %\caption{Plot 1}
    \end{subfigure}
    \hfill
    \begin{subfigure}[b]{0.3\textwidth}
        \centering
        \includegraphics[width=\textwidth]{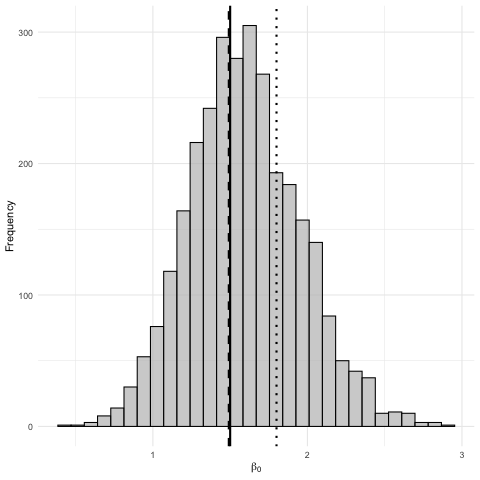}
        %\caption{Plot 2}
    \end{subfigure}
    \hfill
    \begin{subfigure}[b]{0.3\textwidth}
        \centering
        \includegraphics[width=\textwidth]{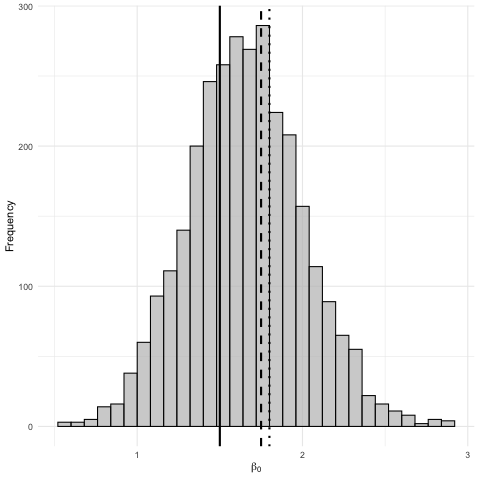}
        %\caption{Plot 3}
    \end{subfigure}
    
    \vspace{0.5cm}  % Add vertical space between rows
    
    % Second row
    \begin{subfigure}[b]{0.3\textwidth}
        \centering
        \includegraphics[width=\textwidth]{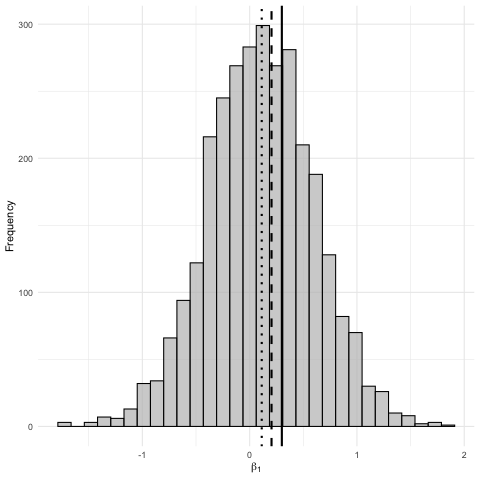}
        %\caption{Plot 4}
    \end{subfigure}
    \hfill
    \begin{subfigure}[b]{0.3\textwidth}
        \centering
        \includegraphics[width=\textwidth]{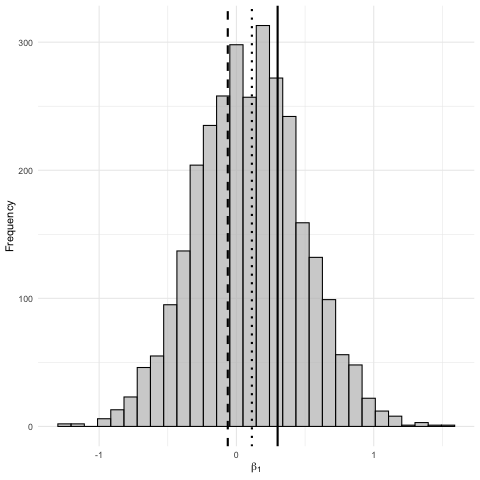}
        %\caption{Plot 5}
    \end{subfigure}
    \hfill
    \begin{subfigure}[b]{0.3\textwidth}
        \centering
        \includegraphics[width=\textwidth]{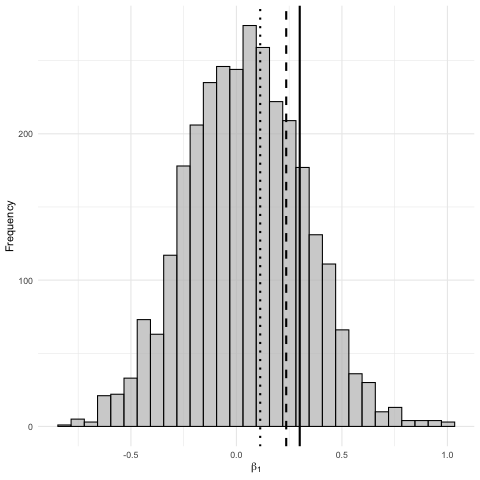}
        %\caption{Plot 6}
    \end{subfigure}
    
    \vspace{0.5cm}
    
    % Third row
    \begin{subfigure}[b]{0.3\textwidth}
        \centering
        \includegraphics[width=\textwidth]{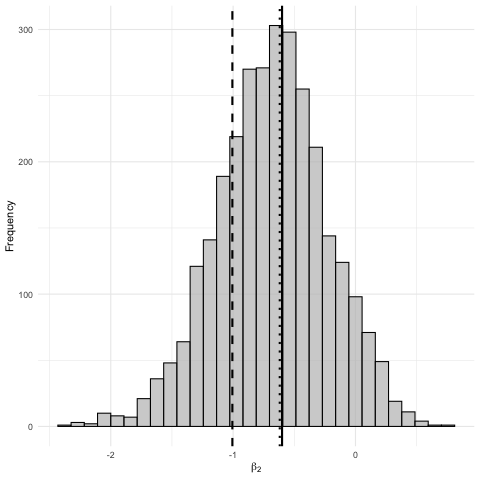}
        %\caption{Plot 7}
    \end{subfigure}
    \hfill
    \begin{subfigure}[b]{0.3\textwidth}
        \centering
        \includegraphics[width=\textwidth]{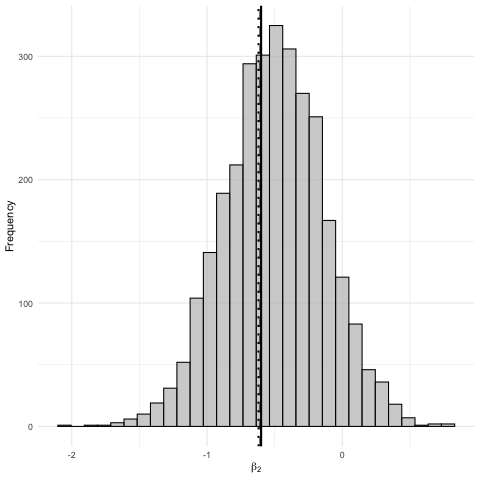}
        %\caption{Plot 8}
    \end{subfigure}
    \hfill
    \begin{subfigure}[b]{0.3\textwidth}
        \centering
        \includegraphics[width=\textwidth]{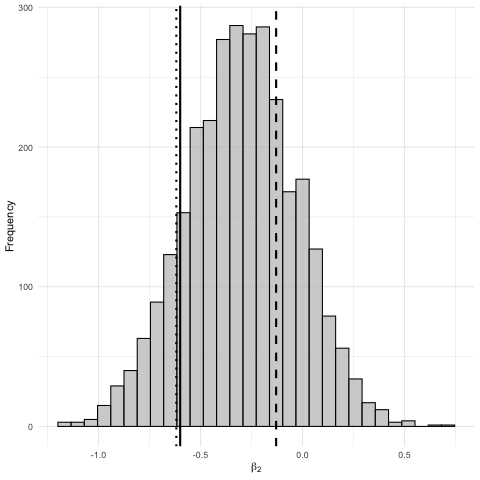}
        %\caption{Plot 9}
    \end{subfigure}
    
    \vspace{0.5cm}
    
    % Fourth row
    \begin{subfigure}[b]{0.3\textwidth}
        \centering
        \includegraphics[width=\textwidth]{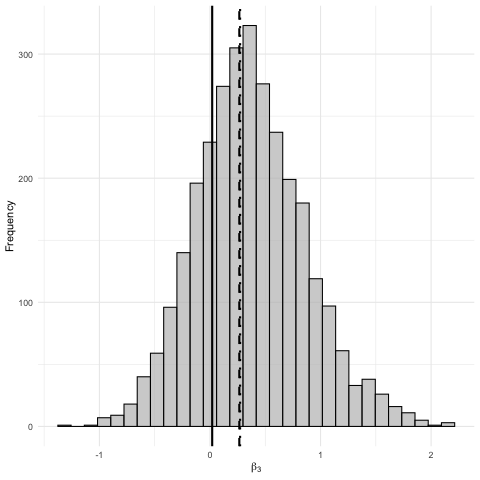}
        %\caption{Plot 10}
    \end{subfigure}
    \hfill
    \begin{subfigure}[b]{0.3\textwidth}
        \centering
        \includegraphics[width=\textwidth]{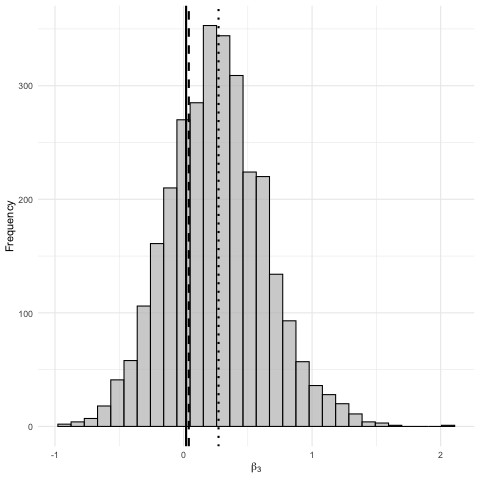}
        %\caption{Plot 11}
    \end{subfigure}
    \hfill
    \begin{subfigure}[b]{0.3\textwidth}
        \centering
        \includegraphics[width=\textwidth]{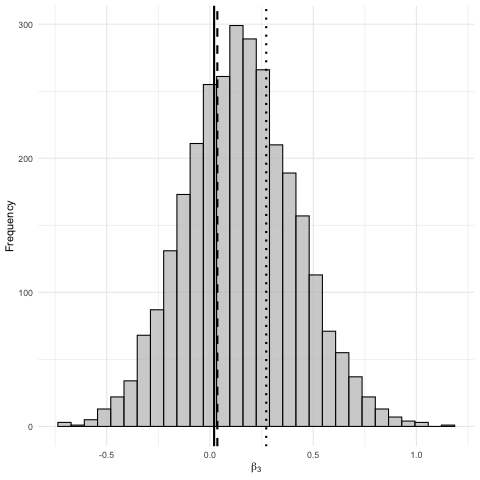}
        %\caption{Plot 12}
    \end{subfigure}
    \caption{Scenarios 2 and 3: Posterior histograms intercept and partial slope coefficients for a single sample of size $n=50$ of a model with $\beta_0=1.5$, $\beta_1=0.3$, $\beta_2=-0.6$ and $\beta_3=0.02$ from top row to bottom row, respectively.   The first column is based on the Vague prior, the second column is based on the Logistic prior with identical priors for intercept and slope coefficients (equations \ref{eq:b0.prior.arbitrary.theta.prior} and \ref{eq:bi.prior.arbitrary.theta.prior}), and the third column is based on the Weighted prior   (Equations \ref{eq:b0.prior.arbitrary.theta.prior.wtd} and \ref{eq:bi.prior.arbitrary.theta.prior.wtd}; see also Section sec:arbitrary.beta).  Each histogram contains three vertical lines showing the true value (solid), the MLE (dotted) and the posterior mean (dashed).}
    \label{F:SingleSample_scenario_2+3cov}
\end{figure}